\documentclass[fleqn,usenatbib]{mnras}

\usepackage{newtxtext,newtxmath}

\usepackage[T1]{fontenc}

\DeclareRobustCommand{\VAN}[3]{#2}
\let\VANthebibliography\thebibliography
\def\thebibliography{\DeclareRobustCommand{\VAN}[3]{##3}\VANthebibliography}

\usepackage{graphicx}	
\usepackage{amsmath}	

\usepackage{amssymb}	
\usepackage{ulem}
\usepackage{comment}
\usepackage{color}
\usepackage{xcolor}
\usepackage{booktabs,caption}
\usepackage[flushleft]{threeparttable}

\newcommand{\HII}{H{\sc ii}\ }
\newcommand{\HI}{H\,\textsc{i}\ }
\newcommand{\kms}{km s$^{-1}$}
\usepackage{tikz,xcolor,hyperref}

\definecolor{lime}{HTML}{A6CE39}
\DeclareRobustCommand{\orcidicon}{%
	\begin{tikzpicture}
	\draw[lime, fill=lime] (0,0) 
	circle [radius=0.16] 
	node[white] {{\fontfamily{qag}\selectfont \tiny ID}};
	\draw[white, fill=white] (-0.0625,0.095) 
	circle [radius=0.007];
	\end{tikzpicture}
	\hspace{-2mm}
}

\foreach \x in {A, ..., Z}{%
	\expandafter\xdef\csname orcid\x\endcsname{\noexpand\href{https://orcid.org/\csname orcidauthor\x\endcsname}{\noexpand\orcidicon}}
}

\title[Sub-kpc scale gas density histogram in M83]{Sub-kpc scale gas density histograms of the nearby barred spiral galaxy M83: Multi-component molecular gas structure reflecting the galactic environment}

\author[R. Matsusaka et al.]{Ren Matsusaka\orcidA{},$^{1}$\thanks{E-mail: ren.matsusaka.jp@gmail.com}
Toshihiro Handa,$^{2,3,4}$
Fumi Egusa\orcidD{},$^{1}$
Yusuke Fujimoto\orcidC{},$^{5}$
Fumiya Maeda\orcidB{},$^{6}$
\newauthor
Takeru Murase\orcidE{},$^{7}$
Yosuke Shibata,$^{2,8}$
Rina Kasai,$^{2}$
Ryo Amano,$^{2}$
Tomoki Ikeda$^{2}$ 
and Tomoki Yamaguchi$^{2}$ 
\\
$^{1}$Institute of Astronomy, Graduate School of Science, The University of Tokyo, 2-21-1 Osawa, Mitaka, Tokyo 181-0015, Japan\\
$^{2}$Department of Physics and Astronomy, Graduate School of Science and Engineering, Kagoshima University, 1-21-35 Korimoto, Kagoshima, \\Kagoshima 890-0065, Japan\\
$^{3}$Amanogawa Galaxy Astronomy Research Center, Kagoshima University, 1-21-35 Korimoto, Kagoshima, Kagoshima 890-0065, Japan\\
$^{4}$Division of Liberal Arts, Kogakuin University, 2665-1 Nakano-cho, Hachioji, Tokyo 192-0015, Japan\\
$^{5}$Department of Computer Science and Engineering, University of Aizu, Tsuruga Ikki-machi, Aizu-Wakamatsu City, Fukushima 965-8580, Japan\\
$^{6}$Research Center for Physics and Mathematics, Osaka Electro-Communication University, 18-8 Hatsucho, Neyagawa, Osaka, 572-8530, Japan\\
$^{7}$Faculty of Engineering, Gifu University, 1-1 Yanagido, Gifu 501-1193, Japan\\
$^{8}$Department of Physics, Nagoya University, Furo-cho, Chikusa-ku, Nagoya, Aichi 464-8601, Japan
}

\date{Accepted 2026 May 14. Received 2026 May 14; in original form 2025 September 24}

\pubyear{\the\year{}}

\begin{document}
\label{firstpage}
\pagerange{\pageref{firstpage}--\pageref{lastpage}}
\maketitle

\begin{abstract}
We investigate the sub-kiloparsec (sub-kpc) molecular ISM structure and its relation to the galactic environment and star formation in the barred spiral galaxy M83 (NGC 5236). We employ the gas density histogram (GDH), which quantifies molecular gas surface density within $550~\mathrm{pc}\times550~\mathrm{pc}\times100~\mathrm{km~s^{-1}}$ cells. The GDHs are well described by one or two log-normal components, corresponding to the lower and higher-surface-density molecular components, referred to as L-LN and H-LN, respectively. The L-LN mass ($M_{\rm L}$) is relatively uniform across the disk, whereas the H-LN mass ($M_{\rm H}$) is highly structured and traces spiral arms. The fractional contribution of the H-LN component ($f^{\prime}_{\rm H}$) shows coherent structures across the disk and is enhanced along spiral arms, consistent with our previous Milky Way results. Moreover, while the L-LN correlates only weakly with star formation rate surface density ($\Sigma_{\rm SFR}$) and shows a steep Kennicutt-Schmidt (KS) relation with surface-density saturation reminiscent of atomic gas, the H-LN exhibits a tighter, nearly linear correlation similar to the conventional molecular KS relation. These results provide direct evidence that the molecular gas in M83 consists of multiple components. Star formation is more closely linked to the H-LN component, whereas the L-LN component appears to represent a more spatially extended molecular gas. Overall, our results suggest that galactic environments control the relative contribution of the two LN components, and that enhanced H-LN contribution is associated with elevated star formation activity.
\end{abstract}



\section{Introduction}
Star formation is driven by the transformation of diffuse interstellar matter (ISM) into dense molecular clouds (MCs). However, the early stages from extended gas to MC formation remain poorly constrained observationally, motivating studies of the ISM across the full density range including MC envelopes and inter-cloud gas. Previous studies in the Milky Way have mainly focused on MC and/or giant molecular cloud (GMC) scale \citep[e.g.][]{MuraseETAL2023,KohnoETAL2024,SchneiderETAL2025}. These MCs typically have masses of $M_{\rm{mol}} \sim 4\times10^5 M_{\odot}$ and sizes of $ \sim 40$ pc \citep[e.g.][]{Scoville&Sanders1987,Roman-DuvalETAL2010,ColomboETAL2022}, serving as key sites for star formation activity in galaxies. 

Studies of nearby galaxies, the advantage of a face-on view has been utilized to reveal that massive GMCs are predominantly located in spiral arms, suggesting that the spiral arms play a crucial role in the physical properties of the GMCs \citep[e.g.][]{EgusaETAL2011,ColomboETAL2014}. More recently, studies with the Atacama Large Millimeter/submillimeter Array (ALMA) have achieved $\lesssim$100 pc resolution in nearby galaxies, enabling detailed investigations of GMCs \citep[e.g. under 50 pc resolution;][]{DemachiETAL2024,KonishiETAL2024,HirotaETAL2024}. In particular, the PHANGS-ALMA survey \citep[Physics at High Angular Resolution in Nearby GalaxieS;][]{LeroyETAL2021} mapped $\sim$70 nearby galaxies within 20 Mpc and greatly expanded the GMC sample \citep[e.g.][]{RosolowskyETAL2021}. Recent ALMA studies have revealed that GMCs share gravitationally bound properties across galaxies but exhibit strong variations in their characteristics and spatial distributions depending on the galactic environment. Although these observations have substantially increased the sample size and improved resolution, they have not yet led to a unified view of how GMC properties are shaped by galactic environment.

These studies have enhanced our understanding of star formation in GMCs. At the same time, observations suggest the presence of an additional low-surface-brightness, spatially extended molecular component. For example, \cite{PetyETAL2013} demonstrated that such extended gas in M51 contributes nearly 50$\%$ of the total detected CO luminosity. Such diffuse contributions can strongly influence observed scaling relations and global trends in galaxy studies \citep{ShettyETAL2014,MaedaETAL2020}. Furthermore, GMC identification methods point to the presence of this component, as roughly half of the CO luminosity is not assigned to identified GMCs \citep{Rosolowsky2007,ColomboETAL2014}. We should note that GMC properties (and thus the amount of gas not assigned to GMCs) depend on the identification methods and parameters therein. Therefore, it is essential to investigate the evolution of the ISM in a framework that goes beyond a dependence on GMC identification.

One underlying cause of the diffuse gas being overlooked is the initial cloud identification stage, which typically relies on fixed beam-size and intensity thresholds. These criteria are tightly connected with the maximum and minimum size of a cloud candidate, which limits the dynamic range of cloud sizes \citep{YanETAL2022}. It is increasingly recognized that imposing such limits is not ideal. In fact, recent numerical simulations suggest that MCs are not isolated and discrete structures but rather a part of a continuous medium \citep{ColmanETAL2024,NiETAL2025}. This has also been confirmed observationally in the Milky Way. To overcome these limitations, some studies \citep[e.g.,][]{SunETAL2018} have employed a statistical approach that characterizes the overall properties of molecular gas without relying on explicit cloud boundaries. In these methods, the observational data were convolved to common spatial and spectral resolutions.

We characterize the density structure of the ISM using the Gas Density Histogram (GDH),  i.e., a normalized histogram of the gas volume (respectively area) fraction as a function of the volume (respectively column) density of the ISM \citep[e.g.][]{EgusaETAL2018,MatsusakaETAL2024}. This approach mitigates biases from cloud identification and captures cloud envelopes and more diffuse components that would otherwise be excluded from a cloud-based analysis. The GDH analysis is constructed directly from observed molecular line cube data without relying on pre-identified MCs. 

In \citet{MatsusakaETAL2024}, sub-kpc-scale GDHs in the Milky Way were shown to be well represented by one or two log-normal (LN) components, suggesting that the molecular-gas distribution can often be decomposed into the lower- and higher-surface-density components, referred to as L-LN and H-LN. The GDH parameters derived from these components exhibit coherent structures on the $l$–$v$ (position-velocity) plane and are closely associated with the locations of classical spiral arms \citep[e.g.,][]{DameETAL2001,ReidETAL2016}. This result suggests that galactic environment systematically controls the relative contribution of the L-LN and H-LN components. However, because the Milky Way is viewed edge-on, it is difficult to spatially separate and unambiguously identify the environmental contexts, such as spiral arms and inter-arm regions. To mitigate the limitations arising from the edge-on orientation of the Milky Way, we apply the GDH method to the nearby face-on galaxy M83 (Table~\ref{table:sample_data}). M83 is a well-known Milky Way analogue \citep[e.g.][]{ChurchwellETAL2009}, and its face-on orientation allows galactic structures to be more clearly resolved and directly compared with GDH characteristics.

We organise this paper as follows. In Section~\ref{sec:data}, we describe the CO and ancillary datasets used in this study. Section~\ref{sec:method} outlines the data selection procedures and the construction of the GDHs. In Section~\ref{sec:results}, we present the GDHs for individual regions within M83 and analyse the spatial distributions of L-LN and H-LN parameters across the galactic disk. 
In Section~\ref{sec:discussion}, we discuss how the relative contribution of the L-LN and H-LN components varies with galactic environment and compare their connections to star formation. We also examine the spatial association of the H-LN fraction with spiral arms and consider possible large-scale processes that enhance the contribution of H-LN structures. Section~\ref{sec:conclusions} presents our conclusions and a summary of our findings.
\begin{figure}
    \includegraphics[width=\linewidth,trim={0 5 15 40}, clip]{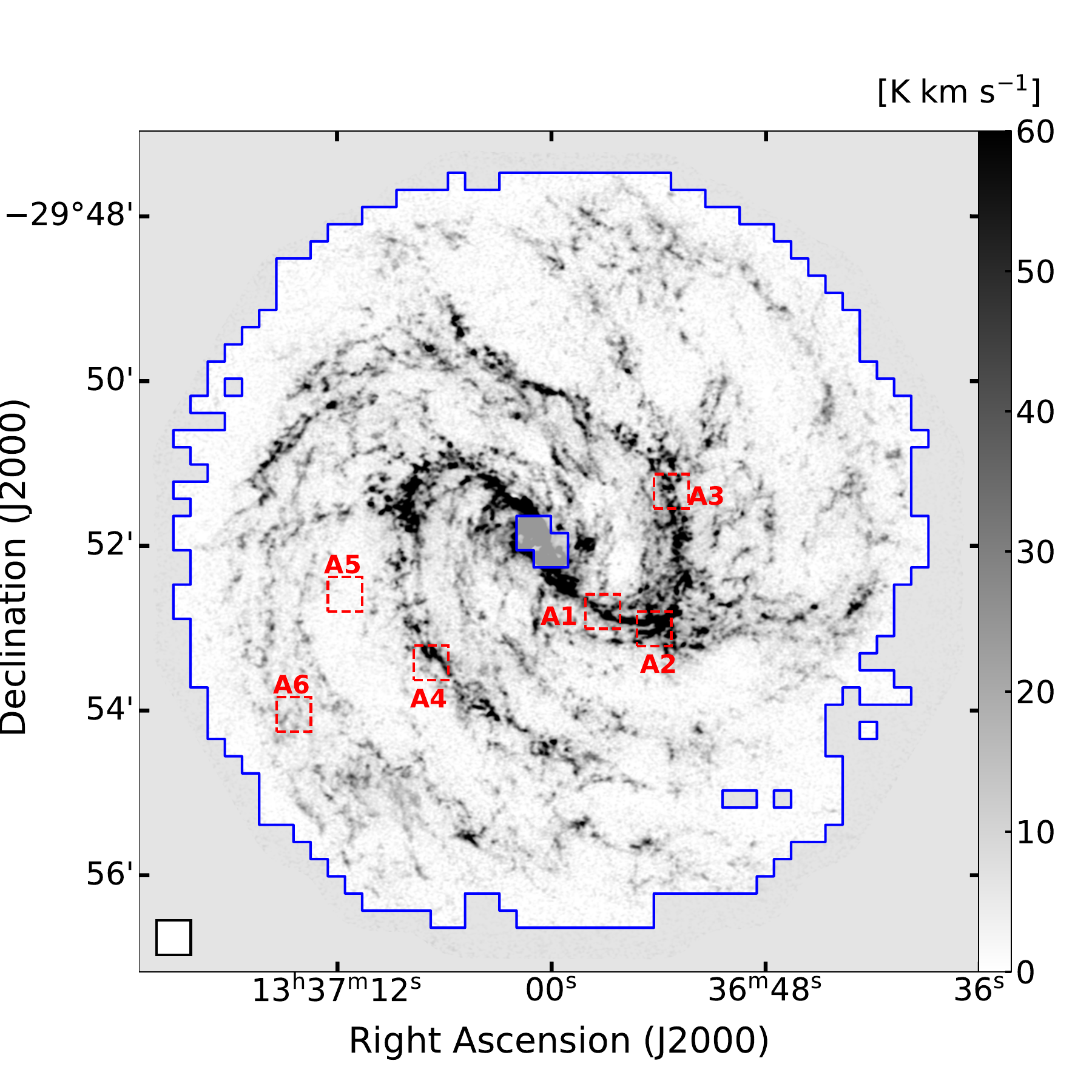}
    \caption{The integrated intensity map of CO($J$ = 1--0) emission of the entire disk of M83 from ALMA 12 m + 7 m + TP observations \citep{KodaETAL2023}. The 550 pc box (100$\times$100 pixel) used to construct each GDH is shown as a square at the bottom-left corner. GDH-cell centres are sampled on a grid with 50 pixel spacing, resulting in a 50\% overlap with adjacent cells along both the horizontal and vertical axes. The blue boundary indicates the footprint of valid GDH cells used in the analysis, i.e. the union of 100$\times$100 pixel GDH cells for which the GDH analysis was successfully performed. Six boxes with red dashed lines are the representative GDH cells shown in Fig.~\ref{fig:M83GDH_sample}.}
    \label{fig:M83image}
\end{figure}

\section{data}
\label{sec:data}

\subsection{CO cube data of the nearby Milky Way-like galaxy M83}
The global and observational properties of the M83 dataset \citep{KodaETAL2023,KodaETAL2025} are summarized in Table~\ref{table:sample_data}. The dataset covers the entire disk of M83, including the outer disk, allowing us to investigate spatial changes in the GDH properties across the disk. With an original mass sensitivity of $\sim 10^{4}~M_{\odot}/\rm{beam}$ (3$\sigma$) and a spatial resolution of 40 pc, the data are sensitive to typical Milky Way MCs and their envelopes ($M_{\rm mol} \sim 4 \times 10^{5}~M_{\odot}$, size $\sim 40$ pc). This high sensitivity is particularly important for our analysis, because the GDHs are intended to characterize the full CO-emitting gas distribution, including low-surface-brightness components such as extended envelopes around MCs. Compared to shallower surveys such as PHANGS-ALMA, this dataset is therefore better suited to recovering faint molecular emission that may significantly contribute to the overall gas distribution.

\begin{table}
    \begin{threeparttable}
    \centering
        \caption{Global properties$\And$observational parameters of M83}
        \label{table:sample_data}
        \begin{tabular}{|l|l|}
            \\ \hline
            Parameter & Value \\ 
            \hline\hline
            Morphological type$^{\rm{a}}$ & SAB(s)c  \\
            Center position (J2000.0)$^{\rm{b}}$ & 13$^{\rm{h}}$37$^{\rm{m}}$0$^{\rm{s}}$.57, -29$^{\circ}$51'56''.9  \\
            Distance (Mpc)$^{\rm{c}}$ &4.5 \\
            Inclination (deg.)$^{\rm{d}}$  &26 $\pm$ 2 \\ 
            Position Angle (deg.)$^{\rm{d}}$ & 225 $\pm$ 1 \\
            Systemic Velocity (km s$^{-1}$)$^{\rm{e}}$ & 510\\
            \hline
            Frequency$^{\rm{f}}$ &115.271204 GHz \\
            Sensitivity (1$\sigma$)$^{\rm{f}}$ & 1.25\ $M_\odot~\rm pc^{-2}$ \\
            Spatial resolution$^{\rm{f}}$ & 40 pc \\
            \hline
        \end{tabular}
        \begin{tablenotes}
          \small    
          \item $^{\rm{a}}$ \cite{deVaucouleursETAL1991book}
          \item $^{\rm{b}}$ \cite{ThatteETAL2000}
          \item $^{\rm{c}}$ \cite{ThimETAL2003}
          \item $^{\rm{d}}$ \cite{KodaETAL2023}
          \item $^{\rm{e}}$ Nobeyama 45 m: \citep{HandaETAL1990}, ALMA: \citep{KodaETAL2023}.
          \item $^{\rm{f}}$ Observed and reduced by \cite{KodaETAL2023}.
        \end{tablenotes}
    \end{threeparttable}
\end{table}

We used the $^{12}$CO($J$=1--0) data observed by ALMA’s 12 m, 7 m, and Total Power (TP) arrays \citep{KodaETAL2023}. These data were calibrated with the Common Astronomy Software Application \citep[CASA;][]{CASATeam2022}, and the TP cube was converted to visibilities using the Total Power to Visibilities package \citep[TP2VIS;][]{KodaETAL2019}. The 12 m, 7 m, and TP data were then jointly imaged with the Multichannel Image Reconstruction, Image Analysis, and Display package \citep[MIRIAD;][]{SaultETAL1995,SaultETAL1996}. Details of these processes are explained in \cite{KodaETAL2023} and \cite{KodaETAL2025}. The data have a spatial resolution of $2\farcs12\times1\farcs71$ at a position angle of $-89^\circ$, corresponding to a physical scale of $\sim$40 pc. The field of view covers a diameter of $9\farcm4$ ($\sim$12.3 kpc), providing extensive spatial coverage for detailed analysis. Each pixel corresponds to $0.25\arcsec\times0.25\arcsec$ ($\sim$5.5 pc $\times$ 5.5 pc), ensuring well-sampled imaging. The velocity resolution is 10 km s$^{-1}$, and the typical sensitivity is 40 mK. This corresponds to a molecular gas surface density sensitivity of $\sim1.25\ M_\odot~\rm pc^{-2}$, enabling the detection of MCs. This sensitivity further allows us to detect the envelopes of MCs. Additionally, the resolution of approximately 40 pc is essential for studying intercloud gas, given the average distance between MCs \citep[$\sim$100 pc;][]{RosolowskyETAL2021}.

\subsection{Ancillary Data}
We used several ancillary datasets to measure star formation rate surface  density, including far-ultraviolet (FUV) and infrared (IR) data. The datasets were obtained from the DustPedia archive \citep{ClarkETAL2018}, where multi-wavelength images of nearby galaxies have been homogeneously processed to provide reliable astrometry, background subtraction, and foreground source removal across a broad spectral range. For our analysis, we used the Galaxy Evolution Explorer (GALEX) FUV and Spitzer 24-$\micron$ intensity maps, smoothed to a common spatial resolution of 7.5\arcsec, and measured the average surface brightness within each area (see Section~\ref{subsec:sub-grid}). 

To determine the velocity range (see Section~\ref{subsec:sub-grid}), we also used high-resolution \HI data obtained from The \HI Nearby Galaxy Survey (THINGS) project \citep{WalterETAL2008}, which provides high-resolution, high-sensitivity 21 cm line observations for 34 nearby galaxies. The \HI data were only used to mask regions of CO emission in our analysis, ensuring that our measurements focus on molecular gas (see Section \ref{subsec:data_mask}). The THINGS observations offer an angular resolution of approximately 7\arcsec and a spectral resolution of 5 \kms, providing detailed information on the atomic hydrogen distribution and kinematics that is crucial for accurate masking of our CO data.

\section{Method}
\label{sec:method}
In this section, we describe the method used to analyse the molecular-gas surface density and to construct the GDHs. The procedure consists of four steps. In Step~1, we define the sub-kpc-scale cells, referred to as GDH cells (Section~\ref{subsec:sub-grid}). In Step~2, we apply masks to select high-quality data (Section~\ref{subsec:data_mask}). In Step~3, we calculate the molecular-gas surface density from the $^{12}$CO($J$=1--0) data cube (Section~\ref{subsec:gdh}). In Step~4, we construct histograms of the molecular-gas surface density across the galactic disk after subtraction of the random-noise contribution (Sections~\ref{subsec:gdh} and \ref{subsec:noise_subtraction}). Steps~3 and 4 follow the procedure described in \citet{MatsusakaETAL2024}.

\subsection{Definition of spatial units and GDH cell}
\label{subsec:sub-grid}
In this paper, we introduce several types of spatial elements. The resolution of the original observed map is set by the beam size, while the pixel size specifies the sampling. A GDH cell refers to a block of pixels used to construct a single GDH.

To construct the GDHs, we first define a square area (surface GDH cell size) corresponding to 550 pc with $\sim$ 5.5 pc pixel size which is equivalent to 100$\times$100 pixels in our data. Along the line-of-sight direction, we select 10 consecutive velocity channels based on the \HI data cube (see more detail, section \ref{subsec:data_mask}). This set of voxels (i.e., 100$\times$100 spatial pixels and 10 velocity channels, total 100,000 voxels) is referred to as a GDH cell, and one GDH is constructed for each such cell. \citet{MatsusakaETAL2024} suggested that at least 1,000 pixels (or voxels) are required to reduce the Poisson noise of each density bin of a GDH. We confirmed that adopting slightly different GDH cell sizes (e.g., 400–600 pc) does not significantly affect the results of this study. We define GDH-cell centres every 50 pixels, so that adjacent GDH cells overlap by 50\% along each axis. Each GDH-derived property is assigned to the location of the corresponding cell centre in the subsequent analysis.

\subsection{Data masking and velocity range definition}
\label{subsec:data_mask}
To exclude low-quality GDHs, we applied a mask to each GDH cell. First, to reduce noise from emission-free channels, we used the \HI data \citep{WalterETAL2008} to define the CO velocity range, following \citet{MuraokaETAL2023}. The \HI cube was convolved to $40\arcsec$ resolution and regridded to match the CO data. We derived a representative velocity, $V_{\rm rep}$, from the \HI moment-1 map and adopted the $^{12}$CO($J$=1--0) emission within $V_{\rm rep}\pm 50~$\kms (10 channels). Here, $V_{\rm rep}$ is taken from the \HI moment-1 map at each spatial pixel, and the $^{12}$CO($J$=1--0) velocity range is defined pixel by pixel as $V_{\rm rep}(x,y)\pm 50$~km~s$^{-1}$. We adopt the same velocity width for all pixels in order to keep the statistical definition of the GDH uniform across the map. Second, after applying the \HI-based velocity mask, we computed the cell-averaged integrated CO intensity and excluded GDH cells with signal-to-noise ratio below 3.

Fig.~\ref{fig:M83image} shows the integrated intensity map obtained from the resulting \HI-masked cube. The blue boundary marks the footprint of valid GDH cells, i.e. the union of 100$\times$100 pixel cells for which the GDH analysis was successfully performed. This map is broadly consistent with that of \citet{KodaETAL2023}, with some scatter at low intensities.

\subsection{Histogram of the molecular gas surface density and GDH}
\label{subsec:gdh}
Compared to cloud-based methods, the GDH analysis has several advantages, such as avoiding biases due to cloud identification and capturing ISM components that would not be identified as clouds. Strictly speaking, the GDH corresponds to the probability distribution function (PDF) of gas densities only if the density fluctuations are statistically stationary. In turbulent media, both volume- and column-density PDFs are often close to LN in numerical simulations \citep[e.g.,][]{PadoanETAL1997,FederrathETAL2008,Federrath&Klessen2013}. At the same time, these studies, together with later observational work, suggest that the PDF (or GDH) shape is not determined by simple turbulence alone. For example, self-gravity can produce a power-law tail on the high-density side of the PDF \citep[e.g.,][]{KritsukETAL2011,KhullarETAL2021}. Observational studies also suggest that the shape of PDF can change from environmental effects, including external compression by nearby OB associations or H\,II regions and large-scale converging or colliding flows \citep[e.g.,][]{TremblinETL2014,SchneiderETAL2022}.

Although the brightness distribution index (BDI) has been used in some studies as an alternative to GDH analysis \citep{SawadaETAL2012a,SawadaETAL2012b,SawadaETAL2018}, it is too simplified to fully capture the detailed density structure of the ISM. 
The GDH is defined for the volume density of the ISM, and \citet{MatsusakaETAL2024} constructed GDHs for each sub-kpc-scale pixel of the ISM in the Milky Way. In face-on galaxies, the line-of-sight depth is not directly constrained. We therefore construct GDHs using molecular-gas surface density (column density), rather than volume density, throughout this work. The comparison is made among GDH cells of fixed size, and modest variations in disk thickness within an individual GDH cell are not expected to affect the separation between the low and high density components (see Section \ref{subsec:estimate_dif-den_gas}).

In this work, we calculate the molecular gas surface density on a channel-by-channel basis rather than integrating over the full velocity range. For a voxel with velocity width $\Delta v$, the surface density is given by
\begin{equation}
\Sigma_{\rm mol}\ [{\rm M_\odot\,pc^{-2}}]
= \alpha_{\rm CO(1-0)}\ T_{\rm B}(v)\ \Delta v\ \cos i ,
\label{eq:sigchan}
\end{equation}
where $T_{\rm B}(v)$ is the brightness temperature, $i$ is the inclination of the galactic disk, and $\alpha_{\rm CO(1-0)}$ is the 
intensity-to-surface-density conversion factor.

In this study, we adopt $\alpha_{\rm CO(1-0)} = 3.14~M_\odot\ \mathrm{(K\,km\,s^{-1}\,pc^{2})^{-1}}$ for M83 \citep{LeeETAL2024}. Although \citet{LeeETAL2024} reported a variation of $\sim$2–5 along the radial direction, such changes have little impact on the GDH analysis because the conversion factor can be considered approximately constant on sub-kpc scale. This assumption is further supported by a study of the Milky Way. \citet{KohnoETAL2024} showed that although the conversion factor varies within individual GMCs, the variation becomes negligible when observed with a beam size of $\sim$40 pc, resulting in nearly uniform values. The little effect by the constant $\alpha_{\mathrm{CO}(J=1-0)}$ assumption is also supported by our previous work, which showed that spatial variations in $\alpha_{\rm CO(J=1-0)}$ introduce only minor shifts along the density axis in GDHs \citep[see][]{MatsusakaETAL2024}.

\subsection{Subtraction of the random-noise contribution}
\label{subsec:noise_subtraction}
To evaluate the GDH shape accurately, especially on the low-surface-density side, it is necessary to account for the contribution of random noise. In the present method, the GDH necessarily includes voxels without significant CO emission (i.e. emission-free voxels). Although these voxels do not contain significant emission, they do exhibit random-noise fluctuations and therefore produce a characteristic feature in the GDH: a straight line with slope 1 in the $\log \Sigma_{\rm mol}$--$\log N$ plane. Following the method introduced by \citet{MatsusakaETAL2024}, we subtract this noise contribution from each GDH.

\begin{figure}
    \includegraphics[width=\linewidth,trim={10 30 -5 10}, clip]{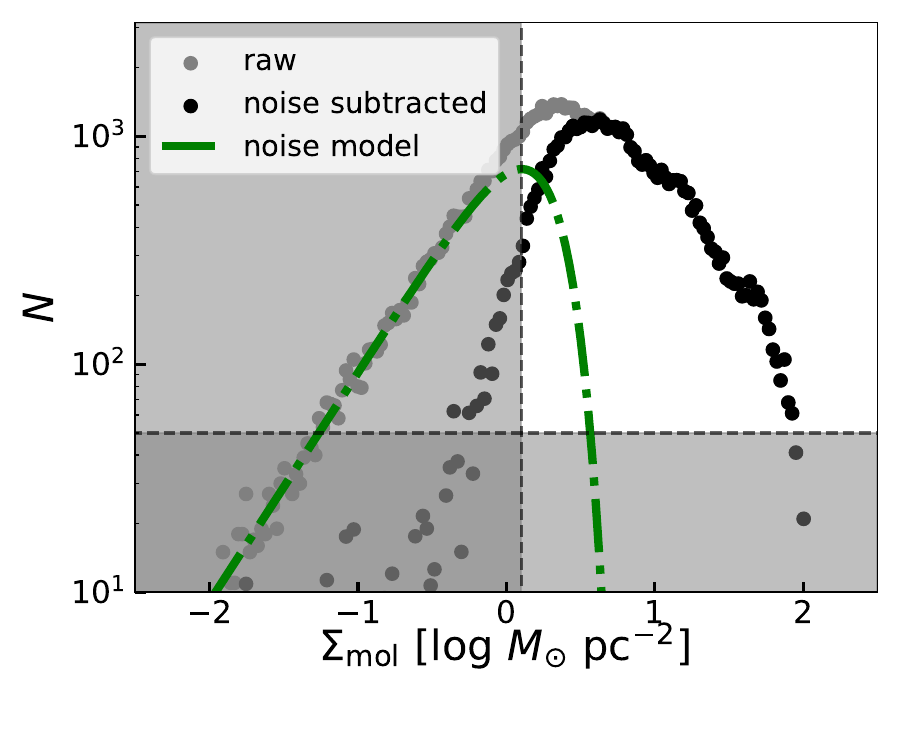}
    \caption{Example of the noise subtraction applied to the GDH. The gray and black points show the raw GDH and the noise-subtracted GDH, respectively. The green dash-dotted line represents the estimated noise model, which accounts for the slope-1 feature in the $\log \Sigma_{\rm mol}$--$\log N$ plane produced by voxels without detected CO signal at low surface densities.}
    \label{fig:noise_subtract}
\end{figure}

Assuming Gaussian noise in $\Sigma_{\rm mol}$, the contribution from voxels without detected CO signal per logarithmic interval can be written as
\begin{equation}
N_{\rm sky}(\Sigma_{\rm mol})
=
A\,\Sigma_{\rm mol}
\frac{1}{\sqrt{2\pi}\sigma_{\rm rms}}
\exp\left(
-\frac{\Sigma_{\rm mol}^{2}}{2\sigma_{\rm rms}^{2}}
\right),
\label{eq:noise}
\end{equation}
where $\sigma_{\rm rms}$ is the rms noise level estimated from emission-free regions, and $A$ is a normalization constant. In the $\log \Sigma_{\rm mol}$--$\log N$ plane, this function asymptotically approaches a straight line with a slope of unity toward low surface densities, reproducing the characteristic slope-1 behaviour produced by voxels without detected CO signal. Fig.~\ref{fig:noise_subtract} compares the raw GDH before noise subtraction with the GDH after subtraction; the green dash-dotted line represents the estimated noise model.

We subtract this component from the observed GDH before fitting the LN components. Although this procedure does not recover the intrinsic GDH below the sensitivity limit, it improves the estimate of the GDH shape near the peak and at the low-density end.

\section{Results}
\label{sec:results}
\begin{figure*}
 \begin{minipage}{0.32\linewidth}
  \centering
  \includegraphics[width=\linewidth,trim={5 5 9 0}, clip, scale=0.333]
  {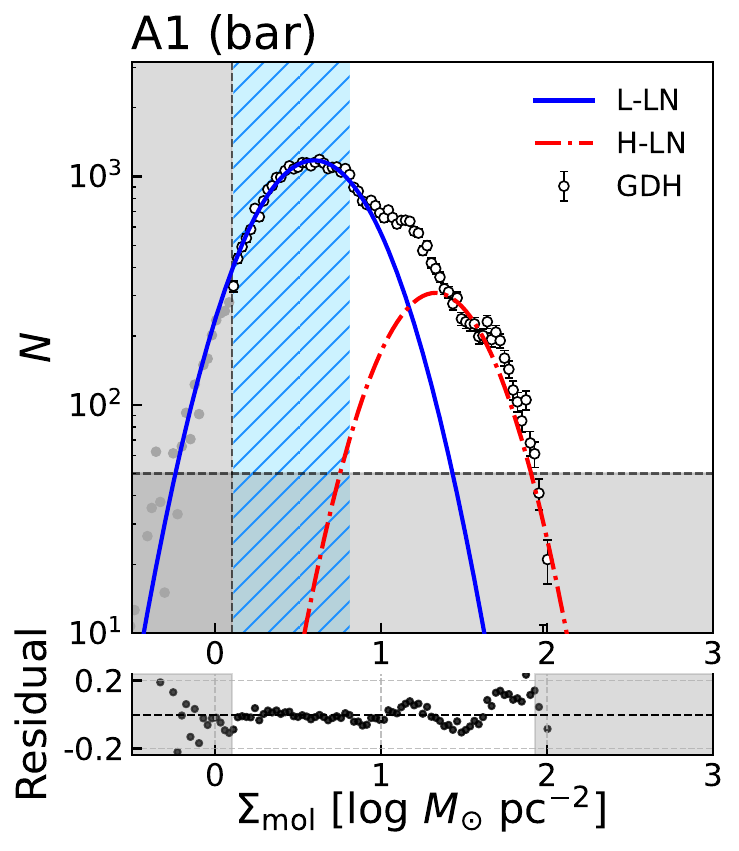}
 \end{minipage}
  \begin{minipage}{0.32\linewidth}
  \centering
  \includegraphics[width=\linewidth,trim={5 5 9 0}, clip, scale=0.333]
  {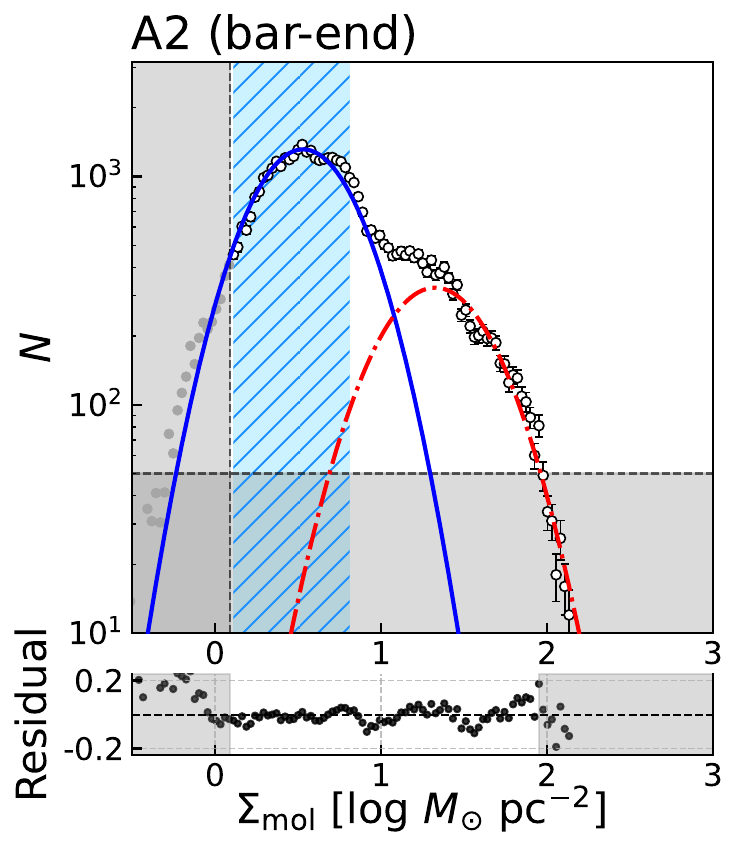}
 \end{minipage}
  \begin{minipage}{0.32\linewidth}
  \centering
  \includegraphics[width=\linewidth,trim={5 5 9 0}, clip, scale=0.333]
  {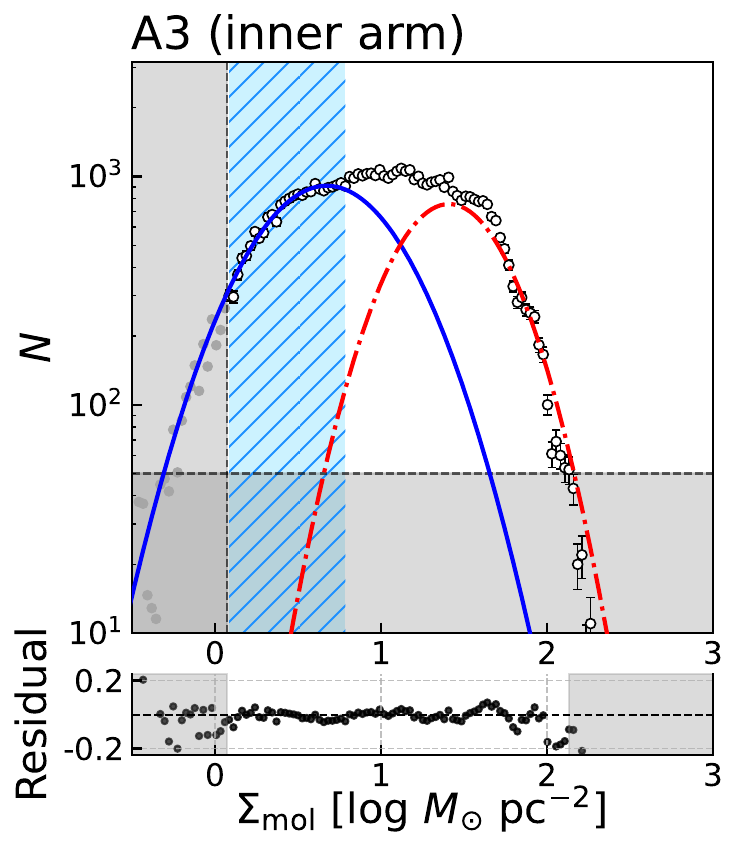}
 \end{minipage}
  \begin{minipage}{0.32\linewidth}
  \centering
  \includegraphics[width=\linewidth,trim={5 5 9 0}, clip, scale=0.333]
  {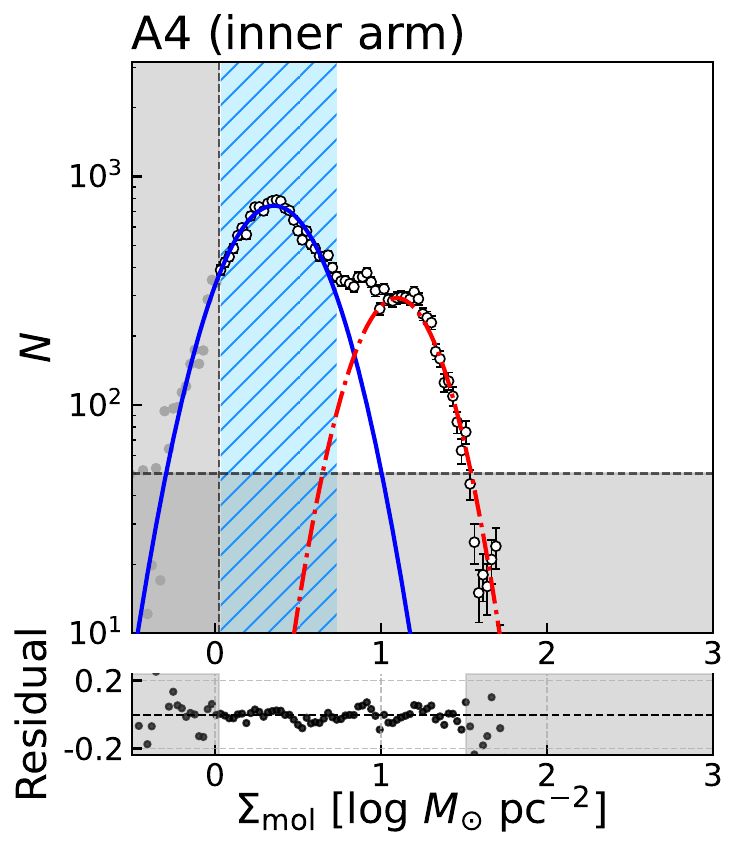}
 \end{minipage}
 \begin{minipage}{0.32\linewidth}
  \centering
  \includegraphics[width=\linewidth,trim={5 5 9 0}, clip, scale=0.333]
  {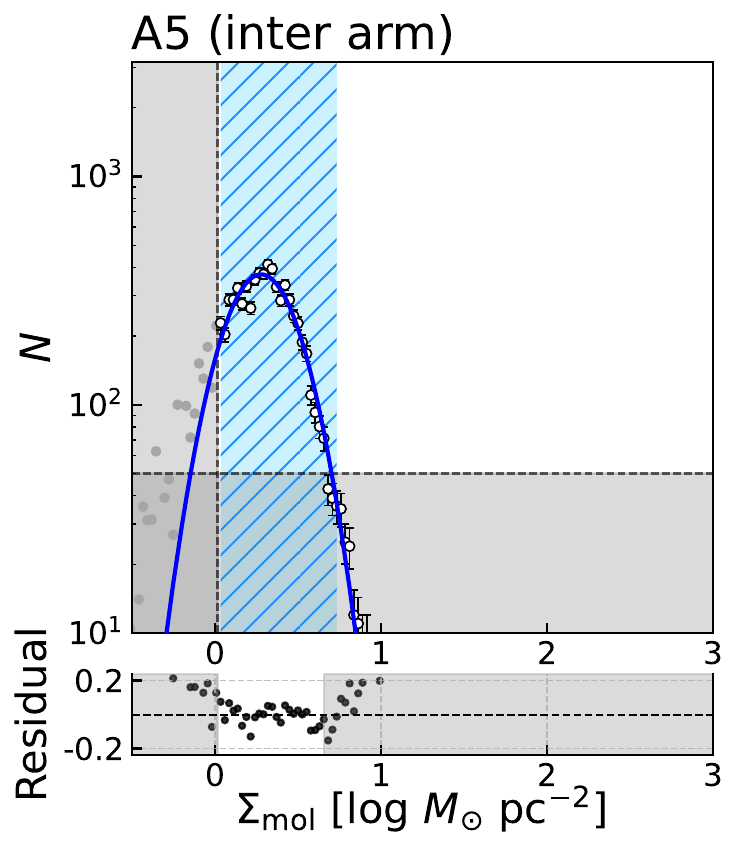}
 \end{minipage}
  \begin{minipage}{0.32\linewidth}
  \centering
  \includegraphics[width=\linewidth,trim={5 5 9 0}, clip, scale=0.333]
  {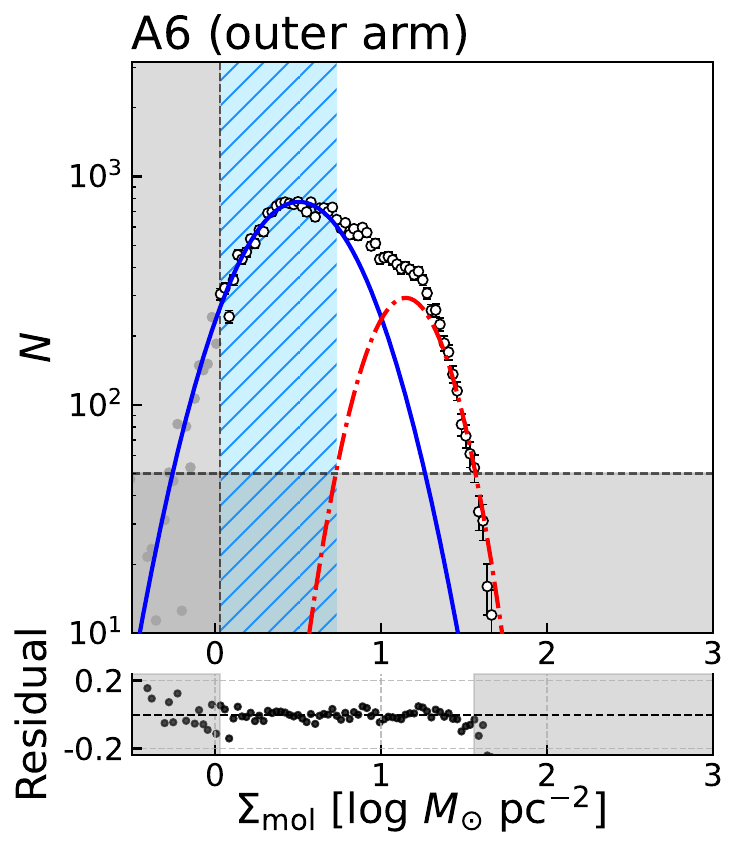}
 \end{minipage}
    \caption{GDHs after removing the random noise component and fitted with two LN components. Each GDH corresponds to areas A1--A6 shown in Fig.~\ref{fig:M83image}. In each GDH, the vertical dashed lines indicate the \(1\sigma\) sensitivity limit, and the horizontal dashed lines indicate the 50 voxels correspond to the number limit. The blue hatched region indicates the fitting range adopted for the L-LN component. The L-LN component is shown by the blue solid line, and the H-LN component by the red dot-dashed line. Data points outside the gray shaded regions were used for the fitting. The absence of an H-LN fit in A5 is due to the insufficient number of voxels per bin on the high-density side. The residuals of the LN fitting, expressed as \(\Delta \log N\) (dex), are shown in the lower panel of each subplot. The all residuals using the fitting (in non-shaded zones) are within the typically $\pm 3 \sigma$.}
    \label{fig:M83GDH_sample}
\end{figure*}
\begin{figure*}
\includegraphics[width=\linewidth,trim={5 5 5 5}, clip]{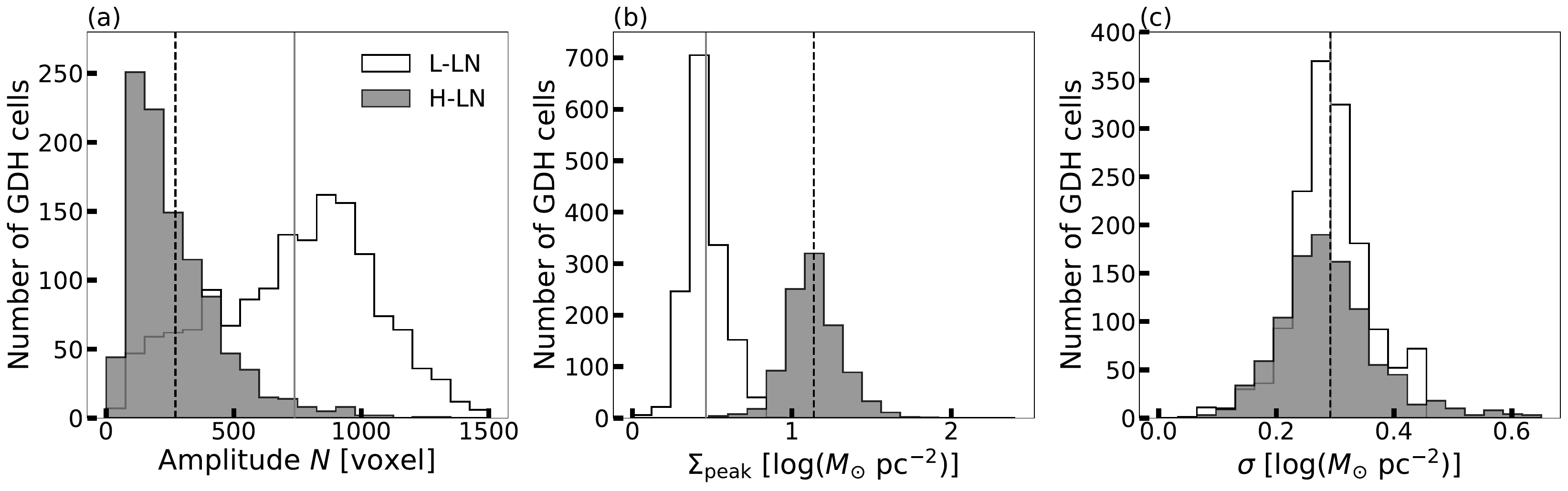}
    \caption{Parameters obtained from the two-component LN fitting, as defined in Equation~(\ref{eq:fitting}). Vertical lines show the average values. Note that the total number of data points for each component differs because, as in Fig. \ref{fig:M83GDH_sample} (panel A5), some regions do not have the H-LN component.}
    \label{fig:fit_param}
\end{figure*}
In this section, we analyze the GDHs derived from $^{12}$CO~($J$=1--0) 
observations of M83 at 40 pc resolution using 550 pc GDH cells. We first present representative GDHs in various galactic environments. We then describe the fitting method used to decompose each GDH into two LN components. Finally, we summarize the statistical properties of these LN components across the entire galaxy and map their spatial distributions.

\subsection{550 pc scale GDH features in M83}
\label{sec:GDH_feature}
Fig.~\ref{fig:M83GDH_sample} shows the GDHs after subtraction of the noise-only component described in Section~\ref{subsec:noise_subtraction}, for six typical shapes of GDHs within M83. The horizontal dashed lines indicate the sufficient voxel number limit based on Poisson statistical uncertainty (50 voxels), while the vertical dashed lines represent the column density limit derived from 1-$\sigma$ noise level in the emission-free channel maps in each GDH cell. Since the column density is primarily derived from the $^{12}$CO~($J$=1--0) emission, the sensitivity limit is governed by the noise level of the $^{12}$CO~($J$=1--0) observations. In this study, a typical sensitivity limit is found at $\Sigma_{\mathrm{mol}} \sim$ 0.35~[$\log (M_{\odot}~ \mathrm{pc}^{-2}$)]. The GDH for less bright regions, such as panel A5, shows a steep decline on the high-surface-density side, consistent with the relative lack of prominent high-surface-density molecular structures in the inter-arm area (see Fig.~\ref{fig:M83image}). In contrast, the other panels exhibit more pronounced high-surface-density features, indicating that such structures contribute more strongly within those GDH cells.

The peaks of these high-surface-density features, around $\Sigma_{\mathrm{mol}} \sim 1$--2 ~[$\log (M_{\odot}~ \mathrm{pc}^{-2})$](see Fig.~\ref{fig:M83GDH_sample} and also Fig.~\ref{fig:fit_param}b), are comparable to those of MCs in the Milky Way, independent of their individual masses or sizes \citep{SolomonETAL1987,Heyer&Dame2015}. However, the peaks on the low-surface-density side are as low as $\Sigma_{\mathrm{mol}}\sim $ 0.5~[$\log (M_{\odot}~ \mathrm{pc}^{-2})$]. Meanwhile $^{12}$CO~($J$=1--0) line is observed from the gas density $n \gtrsim 10^2 \mathrm{cm}^{-3}$, which is smaller than its critical density due to optical thickness. However, it is not inconsistent because the quantities derived from GDHs should not be interpreted as thermodynamical gas densities, since the molecular gas volume filling factor within each beam is expected to be far below unity. Under such circumstances, our estimates represent beam-averaged surface-density structures rather than direct thermodynamical densities, and they may reflect both intrinsic gas properties and the filling factor of molecular structures within the beam \citep{MatsusakaETAL2024}. The L-LN component should be interpreted as a region with a relatively larger contribution from low-surface-density or low-filling-factor molecular structures.

\subsection{L-LN and H-LN components based on GDH analysis}
\label{subsec:estimate_dif-den_gas}
To objectively investigate the shape of the GDH, we parameterize it for each GDH cell. We applied a two-component LN model fitting, as conducted for GDHs in the Milky Way by \citet{MatsusakaETAL2024}. 
Most GDHs were well represented by one or two LN components, expressed as

\begin{equation}
\label{eq:fitting}
\begin{split}
N(\Sigma_{\rm mol}) &= N_{\rm L}(\Sigma_{\rm mol}) + N_{\rm H}(\Sigma_{\rm mol}) \\
&= N_{\rm L} \exp \left[
-\frac{(\log \Sigma_{\rm mol}-\log \Sigma_{\rm peakL})^2}{2\sigma_{\rm L}^2}
\right] \\
&\quad + N_{\rm H} \exp \left[
-\frac{(\log \Sigma_{\rm mol}-\log \Sigma_{\rm peakH})^2}{2\sigma_{\rm H}^2}
\right] .
\end{split}
\end{equation}
where $N_{\rm L}$ and $N_{\rm H}$ are the amplitudes of the L-LN and H-LN components, expressed as voxel counts. The parameters $\Sigma_{\mathrm{peakL}}$ and $\Sigma_{\mathrm{peakH}}$ represent the peak surface densities, while $\sigma$ denotes the width of each LN component. 
\footnote{Note that the fitting was conducted in the $\log \Sigma_{\mathrm{mol}}-N$ domain (where $N$ is the number of voxels) rather than the $\log \Sigma_{\mathrm{mol}}-\log N$ domain, to avoid bias caused by small-number statistics.} The fitting procedure was carried out in two steps. First, we fit the L-LN component around the main low-density peak, using the bins between the sensitivity limit and plus 0.7 dex (blue hatched zone in Fig.~\ref{fig:M83GDH_sample}). Second, we subtracted the fitted L-LN component and then examined the residuals. If the effective number of remaining voxels above the sensitivity limit was 1000 or more, we fitted the H–LN component in addition to the L–LN (Fig.\ref{fig:M83GDH_sample}, panels A1–A4, A6). For GDH cells with fewer remaining voxels, only a single LN component was fitted (panel A5). We carefully examined the residuals from LN fits (see Fig.~\ref{fig:M83GDH_sample}, bottom panel) and found no substantially systematic deviations, particularly around the histogram peaks or the intersection between components. This confirms that the two-component LN model provides a reliable empirical representation of the data.

Although most prior studies fitted GDHs using one or two power-law (PL) distributions \citep[e.g.,][]{BurkhartETALETAL2015,SchneiderETAL2022}, our previous study demonstrated that sub-kpc scale GDHs are better represented by LN components rather than PL distributions \citep{MatsusakaETAL2024}. While PL fits are often interpreted as evidence of gravitationally dominated structures, such an interpretation generally requires sub-parsec spatial resolution \citep[e.g.,][]{KhullarETAL2021,MuraseETAL2023}, which exceeds the capability of our data. We therefore adopt LN functions rather than PL distributions in this work, as our focus is on gas structures at larger scales. Moreover, previous observational studies have suggested that two-LN can reflect multiple physically distinct components whose interpretation depends on environment; examples include ambient turbulent gas plus an externally compressed component, or a lower-column-density atomic/CNM-dominated component plus a higher-column-density H$_2$-dominated component \citep[e.g.,][]{TremblinETL2014,MuraseETAL2023,SchneiderETAL2025}.

\subsection{Characteristics of two LN parameters}
\label{subsec:LN_fit_parameters}
Fig.~\ref{fig:fit_param} presents histograms of the parameters derived from these two LN fits. Panel (a) shows the distributions of the peak number of voxels ($N_{\mathrm{L}}$, $N_{\mathrm{H}}$) for the fitted L-LN (white-filled area outlined with a solid line) and H-LN (gray-filled area) components. The L-LN component generally exhibits a larger peak voxel count (amplitude $N$) than the H-LN component. 

Panel (b) shows the distributions of the peak gas surface densities derived from the LN fits. The L-LN and H-LN components peak ($\Sigma_{\mathrm{peakL}}$ and $\Sigma_{\mathrm{peakH}}$) at approximately $\Sigma_{\mathrm{mol}}$ = 0.45 and 1.1 $[\log (M_{\odot}~\rm pc^{-2})]$, respectively, and this clear difference in peak densities distinctly separates the two LN components. The peak value of the H-LN component is about 5 times higher than that of the L-LN component. This ratio is consistent with the observed trends in the Milky Way, where the typical peak densities of L-LN and H-LN differ by approximately an order of magnitude \citep[e.g.][]{MatsusakaETAL2024}. 

\begin{figure*}
 \begin{minipage}{0.48\linewidth}
  \centering
  \includegraphics[width=\linewidth,trim={8 9 9 5}, clip]
  {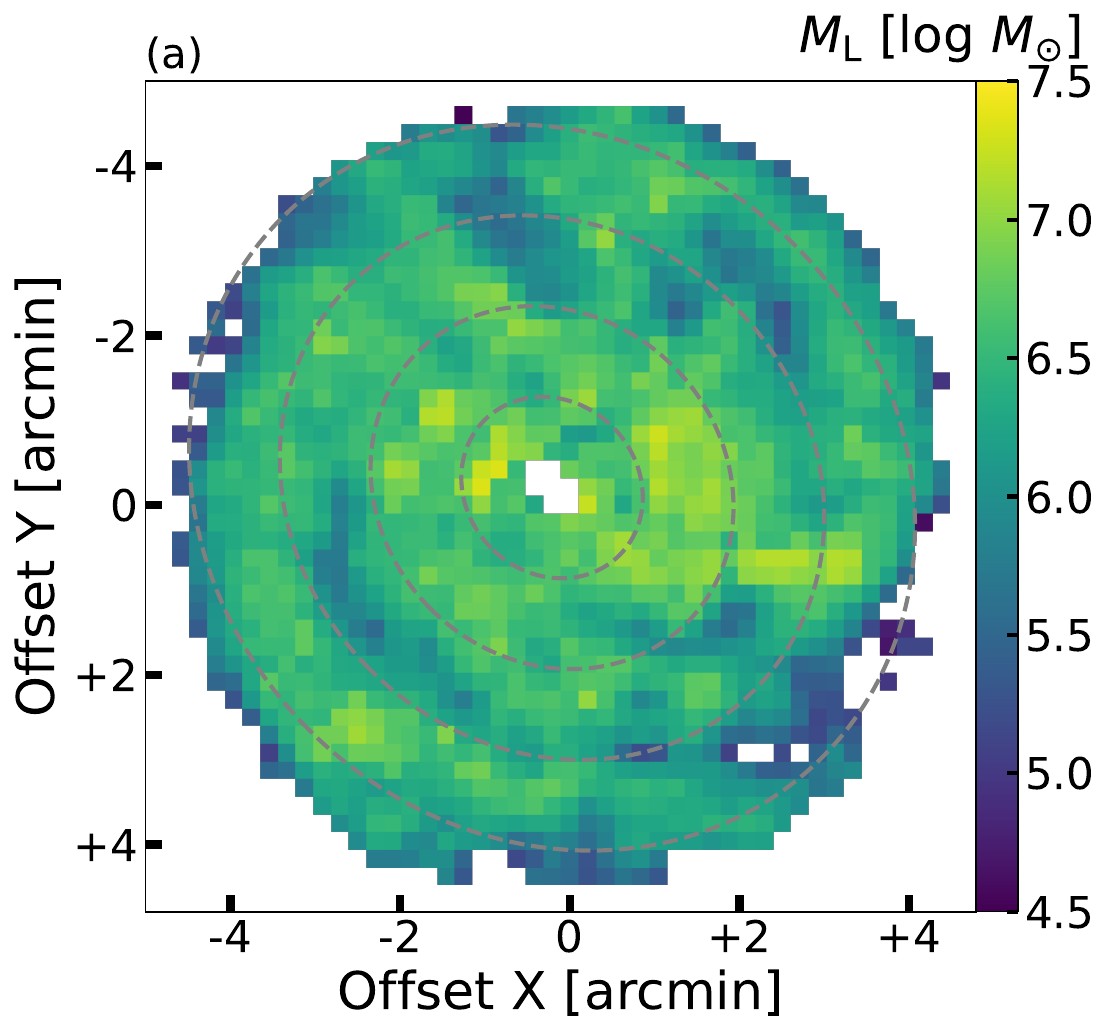}
 \end{minipage}
  \begin{minipage}{0.48\linewidth}
  \centering
  \includegraphics[width=\linewidth,trim={-5 9 9 5}, clip]
  {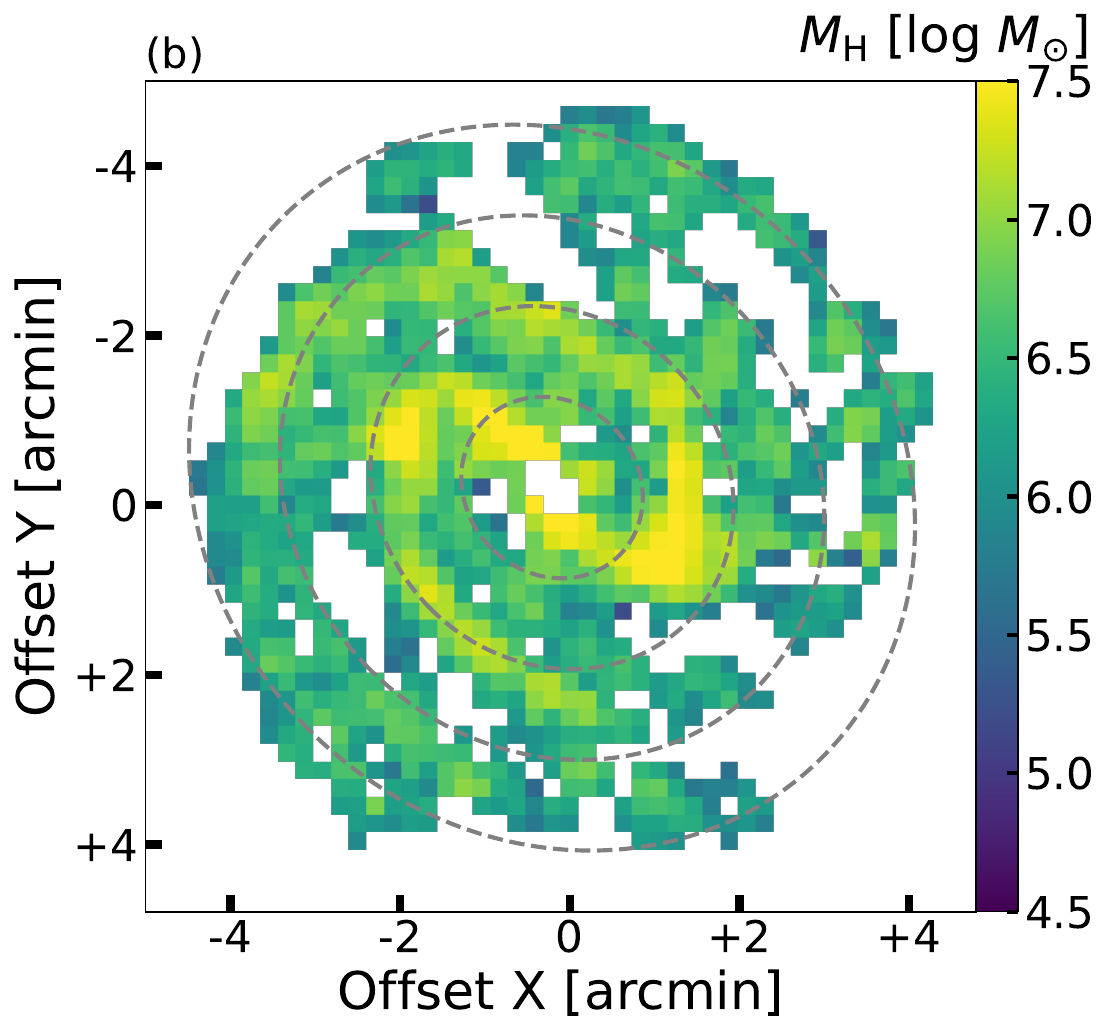}
 \end{minipage}
 \begin{minipage}{0.48\linewidth}
  \centering
  \includegraphics[width=\linewidth,trim={12 25 9 25}, clip]{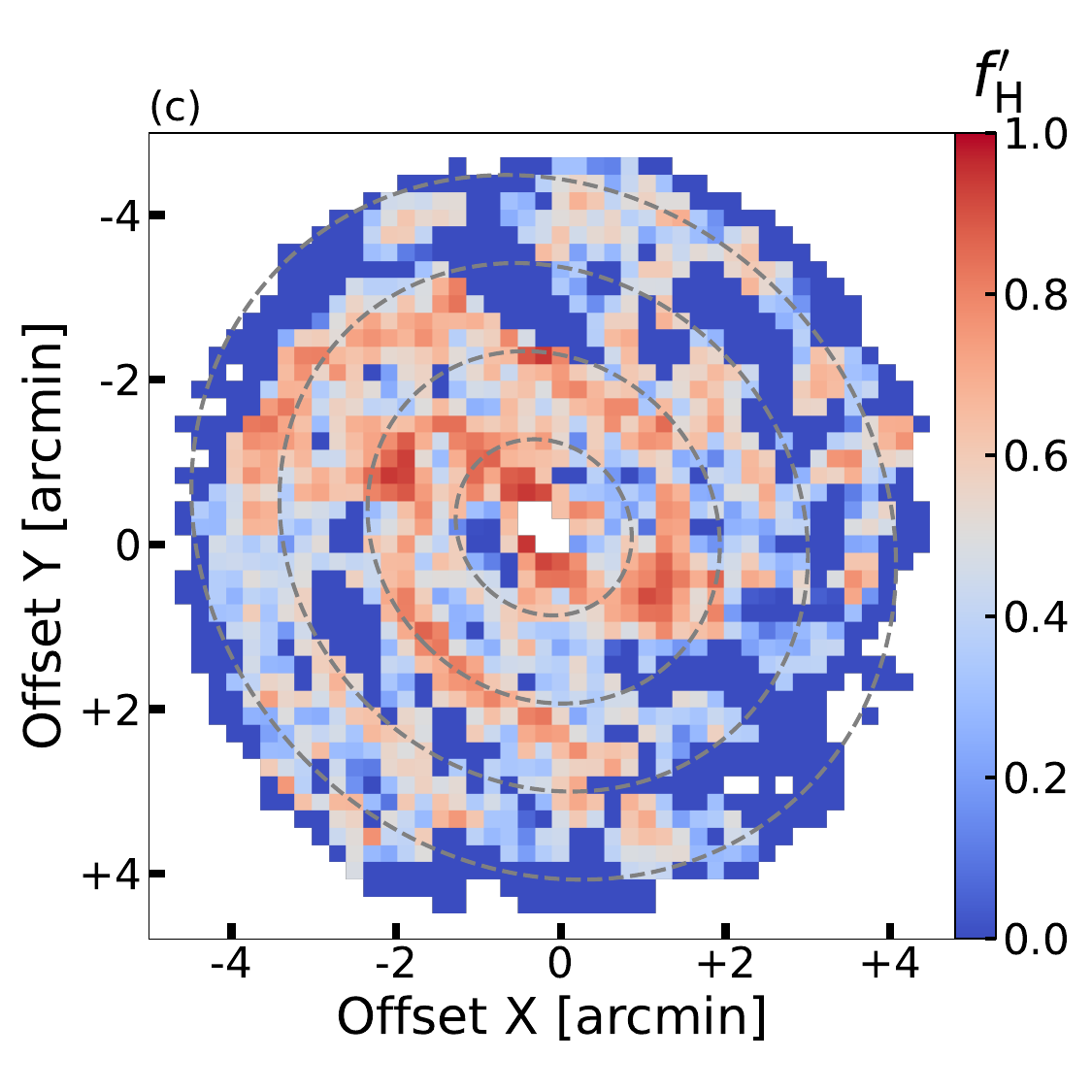}
 \end{minipage}
  \begin{minipage}{0.48\linewidth}
  \centering
  \includegraphics[width=\linewidth,trim={-5 -20 -45 -45}, clip]{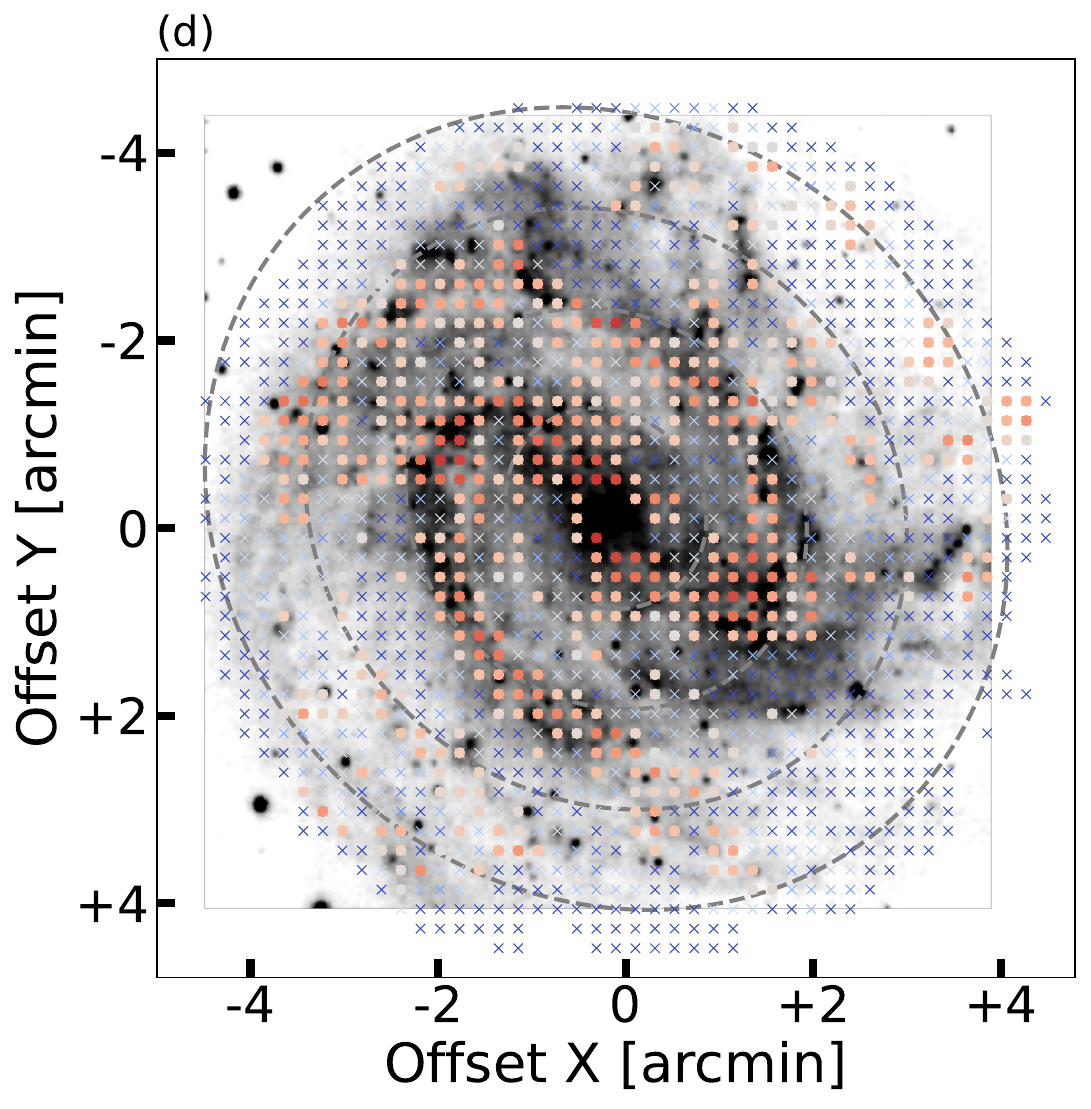}
 \end{minipage}
    \caption{Panel (a) shows the spatial distribution of the total molecular gas mass of the L-LN component ($M_{\rm{L}}$). Panel (b) shows the spatial distribution of the total molecular gas mass of the H-LN component ($M_{\rm{H}}$). In panel (a), blank pixels denote regions masked according to the criteria described in Section~\ref{subsec:data_mask},  while in panel (b), additional blank pixels mark regions where the H-LN component is not detected. (c) The spatial distribution of the $f_{\rm{H}}^{\prime}$ with a resolution of 550 pc based on GDH analysis. Note that $f_{\rm{H}}^{\prime}$ = 0 indicates the regions where the H-LN component could not be detected. Panel (d) shows the Digital Sky Survey (DSS) optical image overlaid with colour markers. The colour markers indicate the centres of GDH cells and are colour-coded by $f_{\rm H}^{\prime}$ using the same colour scale as panel (c). For clarity, cells with $f_{\rm H}^{\prime}<0.5$ are shown as crosses, whereas those with $f_{\rm H}^{\prime}\geq0.5$ are highlighted with circles. The gray dashed circles indicate galactocentric distances, as shown in Fig.\ref{fig:radial_distribution}.}
\label{fig:spatial_dist}
\end{figure*}

Finally, panel (c) shows the widths ($\sigma$) of each LN fit. In the Milky Way, \citet{MuraseETAL2023} reported that the H-LN component has smaller $\sigma$ values than the L-LN component, based on GDHs constructed at a GDH cell size of $\sim$10--50 pc. This implies that the higher-density component occupies a smaller volume fraction with lower turbulence velocities. In contrast, in our sub-kpc scale analysis of M83, the widths of the L-LN and H-LN components are comparable and both concentrated in a narrow range, typically $\sigma\sim$0.2--0.4~$[\log (M_{\odot}~\rm pc^{-2})]$. This indicates that the two components exhibit similar dispersions at  sub-kpc scale.

\subsection{Spatial distributions of L-LN and H-LN}
\label{subsec:spatial_dist}
To quantify the amount of molecular gas represented by each LN component within a GDH cell, we define the molecular-gas masses of the L-LN and H-LN components as

\begin{equation}
\begin{split}
M_{\rm L}\ [M_{\odot}] &= \frac{L^2}{\cos i}\int \Sigma_{\rm mol}\, N_{\rm L}(\Sigma_{\rm mol})\, d \Sigma_{\rm mol},\\
M_{\rm H}\ [M_{\odot}] &= \frac{L^2}{\cos i}\int \Sigma_{\rm mol}\, N_{\rm H}(\Sigma_{\rm mol})\, d\Sigma_{\rm mol},
\end{split}
\end{equation}
where $L^2$ is the projected size of a pixel in pc$^2$ units. To summarize the relative contribution of the two LN components, we define the H-LN fraction as
\begin{equation}
f_{\rm H}^{\prime} = \frac{M_{\rm H}}{M_{\rm L} + M_{\rm H}} .
\end{equation}
Here, $f_{\rm H}^{\prime}$ is defined from the masses of the two fitted components and therefore represents the mass-weighted contribution of the H-LN component within a GDH cell. This definition differs from the area-fraction-based definition ($f_{\rm H}$) adopted in  \citet{MatsusakaETAL2024}. We adopt this form because a mass-fraction-based measure is more commonly used. The calculation using the same definition as in \citet{MatsusakaETAL2024} is presented in Appendix~\ref{app:HLN-fraction}.

\begin{figure}
    \includegraphics[width=\linewidth,trim={5 5 0 5}, clip]{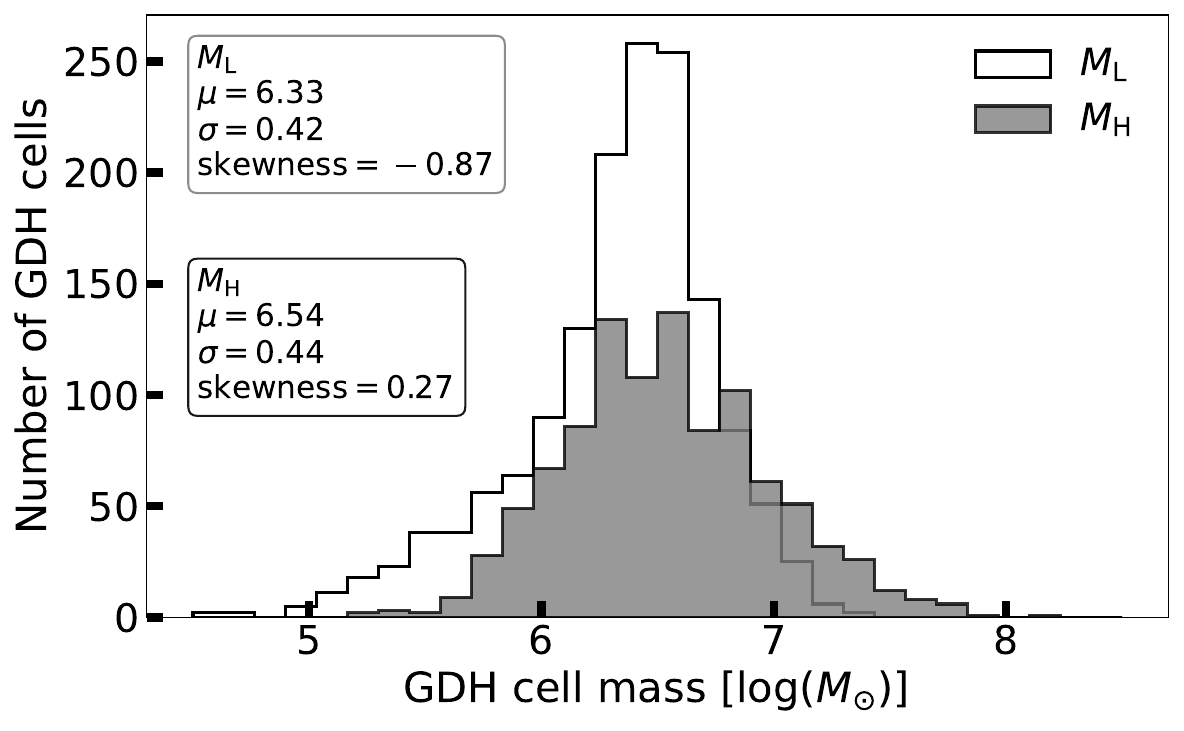}
    \caption{The distributions of GDH cell masses are shown for the L-LN component ($M_{\rm L}$; white) and the H-LN component ($M_{\rm H}$; gray). Here, $\mu$ and $\sigma$ denote the mean and standard deviation, respectively, of the $\log$-transformed masses. The skewness are also derived in logarithmic space.}
    \label{fig:histogram}
\end{figure}

The spatial and statistical distributions of the L-LN and H-LN components (Figs.~\ref{fig:spatial_dist} and \ref{fig:histogram}) reveal clear differences between the two components. While the standard deviations of $M_{\rm L}$ (0.42) and $M_{\rm H}$ (0.44) in logarithmic scale are comparable, their skewness values (-0.87 for $M_{\rm L}$ and 0.27 for $M_{\rm H}$) indicate that the two distributions differ in shape. $M_{\rm L}$ is more sharply peaked around the central value, with a more prominent tail toward the lower-mass side, whereas $M_{\rm H}$ shows a broader and flatter distribution, extending toward the higher-mass side. $M_{\rm{H}}$ tends to concentrate in specific environments, showing a distribution similar to that of dense gas observed in other galaxies using high critical density tracers such as HCN \citep[e.g.,][]{JimETAL2019,NeumannETAL2023}. These findings support the interpretation that the ISM is composed of multiple LN components with distinct spatial and statistical properties, even within the molecular gas traced by CO. This reinforces the idea that star formation models should consider not only the total molecular-gas content but also the relative contribution of different molecular-gas structures within a GDH cell.

Fig.~\ref{fig:spatial_dist}c and d show the spatial distribution of $f_{\rm H}^{\prime}$ across the disk of M83, where $f_{\rm H}^{\prime}$ spans a range from $\sim 0.2$ up to $\sim 0.9$ and becomes $0$ where the H-LN component is not detected. Such GDH cells were included in the calculation of the mean $f_{\rm H}^{\prime}$ by assigning $f_{\rm H}^{\prime}=0$. The mean of $f_{\rm H}^{\prime}$ is 0.57, implying that $\sim$40\% of the molecular-gas mass is contained in the L-LN component. Regions with high $f_{\rm H}^{\prime}$ form coherent structures in the inner disk, whereas the outer disk shows a more fragmented pattern. As shown in Fig.~\ref{fig:spatial_dist}c and d, these high-$f_{\rm H}^{\prime}$ regions are preferentially located along the optically identified spiral arms \footnote{Optical image obtained from the SkyView web page \citep{McGlynnETAL1998}}. This trend is broadly consistent with our previous Milky Way GDH study \citep{MatsusakaETAL2024}, which also suggested a connection between GDH-based parameters and galactic environment, although the nearly face-on view of M83 allows a cleaner environmental interpretation.

\subsection{Radial variation of L-LN and H-LN}
\begin{figure}
    \includegraphics[width=\linewidth,trim={5 5 5 10}, clip]{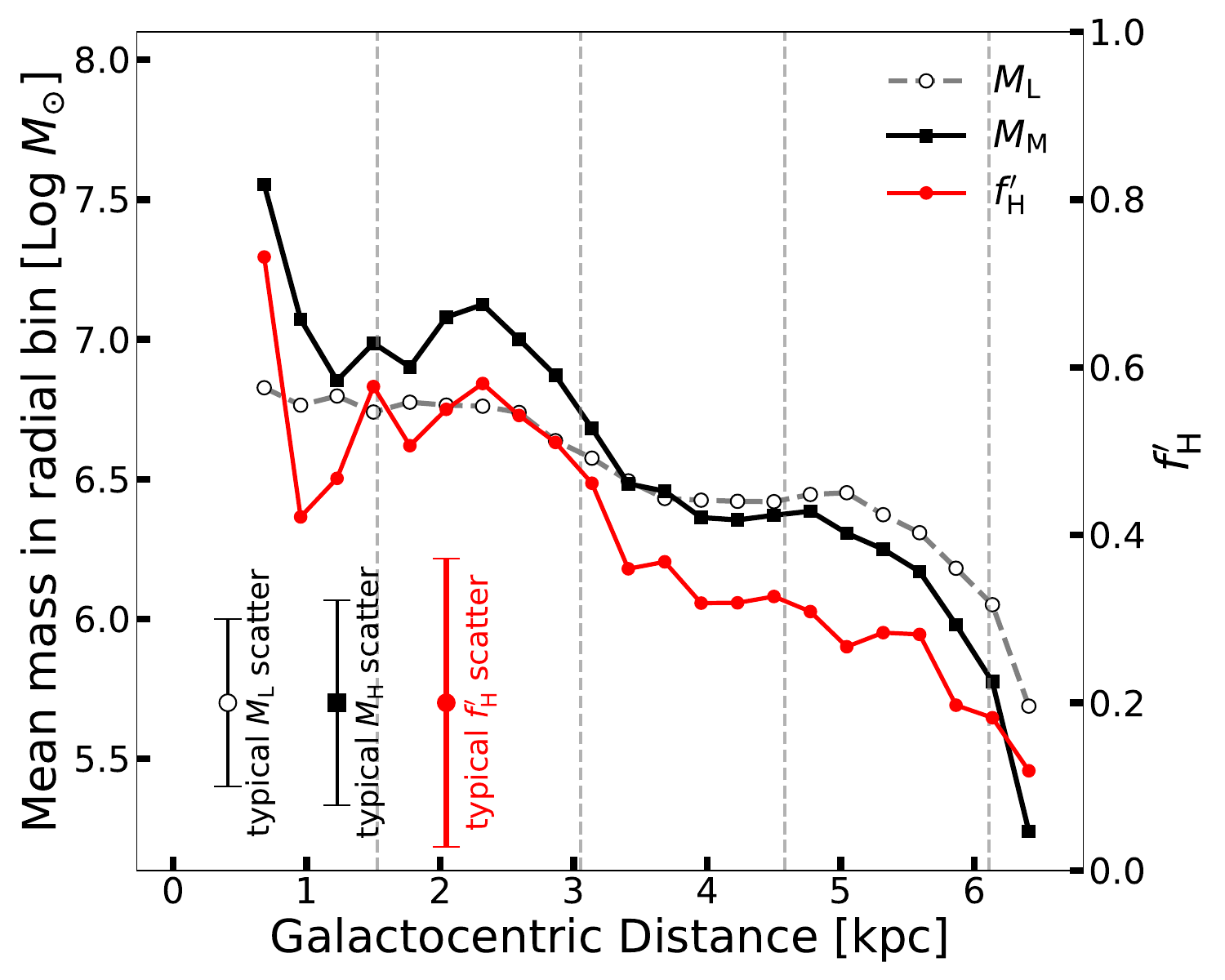}
    \caption{Radial profiles of the average $M_{\rm L}$ and $M_{\rm H}$, shown as white circles connected by a gray dashed line and black squares connected by a black solid line, respectively. The red points and red line show the radial profile of $f'_{\rm H}$, with values given on the right-hand axis. Typical scatter values, quantified using the median absolute deviation, are shown in the lower left. The vertical dashed lines mark the radii corresponding to the gray circles in Fig.~\ref{fig:spatial_dist}.}
\label{fig:radial_distribution}
\end{figure}

Because the most significant difference of galactic environment is the galactocentric radius, we examine how the contributions of the L-LN and H-LN components vary as a function of galactocentric radius (Fig.~\ref{fig:radial_distribution}). In constructing the radial profiles, the average values of $M_{\rm L}$ and $M_{\rm H}$ were calculated separately in each radial bin. For $M_{\rm H}$, GDH cells with no detected H-LN component ($M_{\rm H}=0$) were excluded from the average, so that the profile reflects only cells in which the H-LN component is identified. By contrast, the radial profile of $f'_{\rm H}$ was calculated using all GDH cells, assigning $f'_{\rm H}=0$ to cells without a detected H-LN component.

$M_{\rm L}$ and $M_{\rm H}$ generally decrease from the galactic centre toward the disk edge, with local enhancements in the central region ($<1.5$ kpc), around the bar end ($\sim2$ kpc), and in the outer spiral arm ($\sim4.5$ kpc). In particular, $M_{\rm H}$ is more strongly concentrated toward the centre than $M_{\rm L}$, as is often seen for the total molecular-gas distribution in barred spiral galaxies \citep[e.g.][]{HandaETAL1990,QuerejetaETAL2021}. The central $\sim1$ kpc contains a large amount of ISM and accounts for approximately 15$\%$ of the total molecular gas mass in M83 \citep{KodaETAL2023}, and such centrally concentrated and galactic-environment-associated features are more prominently observed in the H-LN component than in the L-LN. The radial profile of $f'_{\rm H}$ also varies systematically with radius, enabling a direct comparison between the H-LN fraction and the radial behaviour of the two LN components. The radial decline of $f'_{\rm H}$ may therefore reflect not only a change in the relative contribution of the two LN components, but also a transition in the large-scale dynamical environment across M83.

\section{Discussion}
\label{sec:discussion}
In this section, we examine how the H-LN component is related to large-scale galactic environment and how both LN components connect to star formation on sub-kpc scales in M83.

\subsection{Large-scale dynamical effects on the H-LN fraction}
\label{subsec:fden}

The results in Section~\ref{subsec:spatial_dist} show that regions with high $f_{\rm H}^{\prime}$ are not randomly distributed across M83, but instead form coherent structures. They are located along the major spiral arms and around the bar end (Fig.~\ref{fig:spatial_dist}c,d). This spatial correspondence suggests that H-LN becomes more prominent in specific galactic environments, rather than through stochastic local variations alone.

We interpret this trend as evidence that the large-scale dynamical environment in M83 enhances the relative contribution of the H-LN component. In particular, spiral-arm and bar-end regions are likely to promote the compression, accumulation, or interaction of ISM \citep[e.g.][]{KodaETAL2009,QuerejetaETAL2021,QuerejetaETAL2025}. Such environmental effects may increase the contribution of high-surface-density gas within a GDH cell either by changing the physical conditions of the molecular gas, such as its density and temperature, or by increasing the filling factor of dense molecular structures within the beam. In this sense, the high $f_{\rm H}^{\prime}$ values observed along the spiral arms indicate that H-LN becomes dominant where galactic-scale dynamics efficiently shift the molecular-gas distribution toward higher-surface-density structures.

A related result was reported by \citet{KodaETAL2025}. They found that the CO($J$=2--1)/CO($J$=1--0) ratio ($R_{21}$) is enhanced along the spiral arms of M83. This suggests that filling-factor variations alone may not fully explain the arm/inter-arm contrast and that changes in the physical conditions of the molecular gas may also be involved. Taken together, these results suggest that dynamically organised environments shift the molecular-gas distribution toward a more H-LN-dominated state, although the present analysis does not by itself uniquely distinguish changes in intrinsic density from changes in filling factor.

This interpretation is also supported by the large-scale radial change in the molecular-gas morphology of M83 (Fig.~\ref{fig:radial_distribution}). High-sensitivity CO imaging has shown that the inner disk is characterized by spatially coherent molecular spiral arms, whereas the outer disk becomes less coherent and more flocculent \citep[][and our results in Fig.~\ref{fig:spatial_dist}c,d]{KodaETAL2023}. Such a structural transition is qualitatively consistent with theoretical pictures in which the inner disk is more strongly influenced by the bar, while the outer spiral structure is more transient and dynamically evolving \citep{Baba2015}. Although our data do not directly identify the mechanism responsible for the L-LN and H-LN balance, these results suggest that the large-scale dynamics of M83 may help regulate how strongly the H-LN component is enhanced relative to the L-LN component.

We do not attempt to identify the detailed physical mechanism responsible for this environmental dependence. Classical spiral shocks, cloud--cloud interactions, and other large-scale processes may all contribute, but distinguishing among them is beyond the scope of the present data. For example, if the H-LN enhancement is related to a density-wave-like response of the molecular gas, azimuthal offsets between gas structures and recent star formation tracers may be expected \citep[e.g.,][]{EgusaETAL2004,EgusaETAL2009}. Our effective GDH-cell size of $\sim$550 pc, however, is too coarse to test such $\sim$100 pc-scale predictions directly. Local feedback from recent star formation may also influence the molecular-gas structure \citep{FujimotoETAL2020,FujimotoETAL2020b}, but \citet{EgusaETAL2018} showed that its effect on GDH analysis is likely small with the current resolution, since masking regions around \HII regions and supernova remnants does not significantly alter the resulting GDHs. A more direct connection between GDH-based quantities and the underlying dynamical processes will require spatial resolution improved by a factor of several.

\subsection{Sub-kpc scale star formation relation in M83}
\label{subsec:sfr}
In this section, we examine how the star formation rate surface density ($\Sigma_{\rm SFR}$) correlates with molecular-gas surface density on sub-kpc scales. For this purpose, we express the contributions of the L-LN and H-LN components as GDH-cell-averaged surface densities, so that they can be directly compared with $\Sigma_{\rm SFR}$ and with the conventional molecular Kennicutt--Schmidt (KS) relation  \citep{Schmidt1959,Kennicutt1998}. 
\begin{figure}
    \includegraphics[width=\linewidth,trim={-5 5 -5 5}, clip]{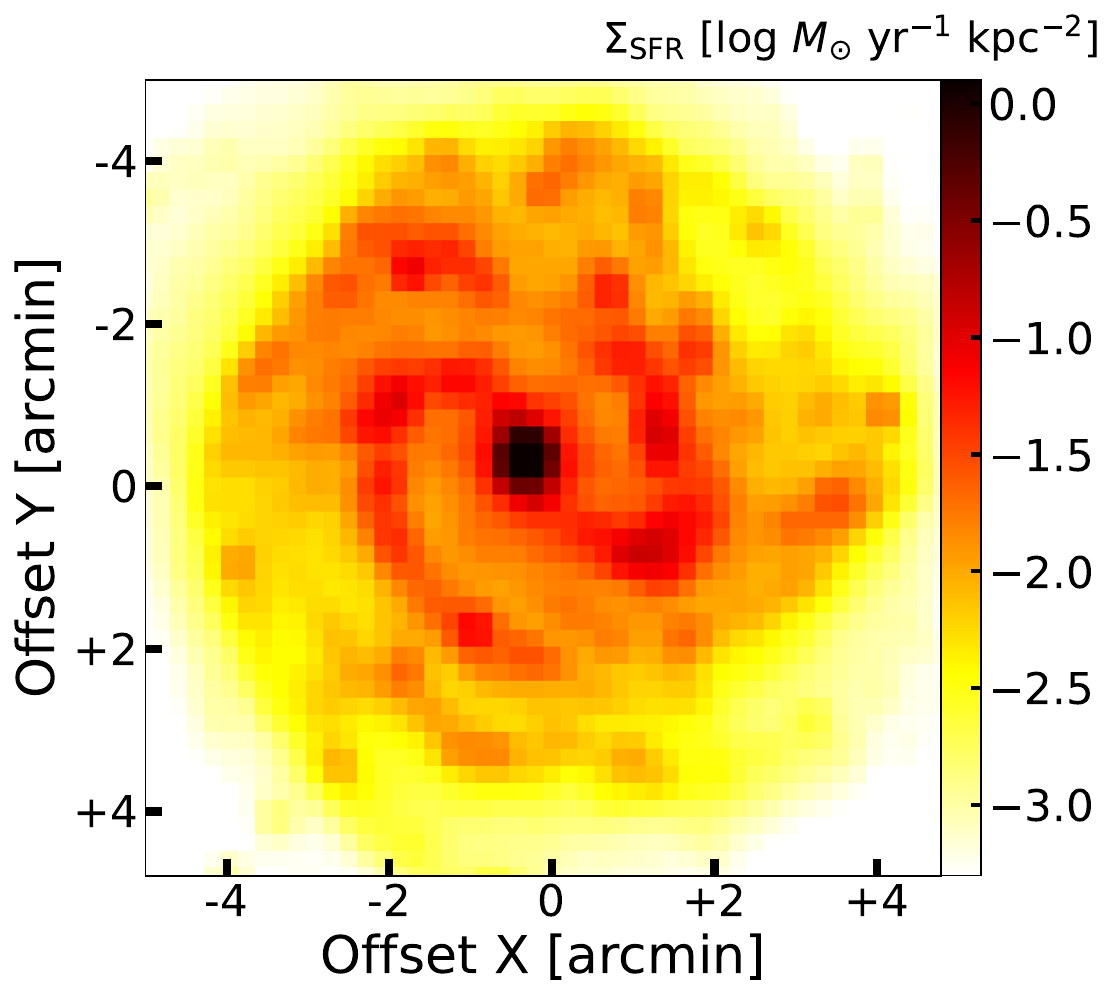}
    \caption{Spatial distribution of $\Sigma_{\rm SFR}$ across M83. The $\Sigma_{\rm SFR}$ values are derived for each GDH cell from the combination of the GALEX FUV and Spitzer 24~$\micron$ intensities.}
    \label{fig:SFR-map}
\end{figure}
\begin{figure*}
    \includegraphics[width=\linewidth,trim={150 0 200 60}, clip]{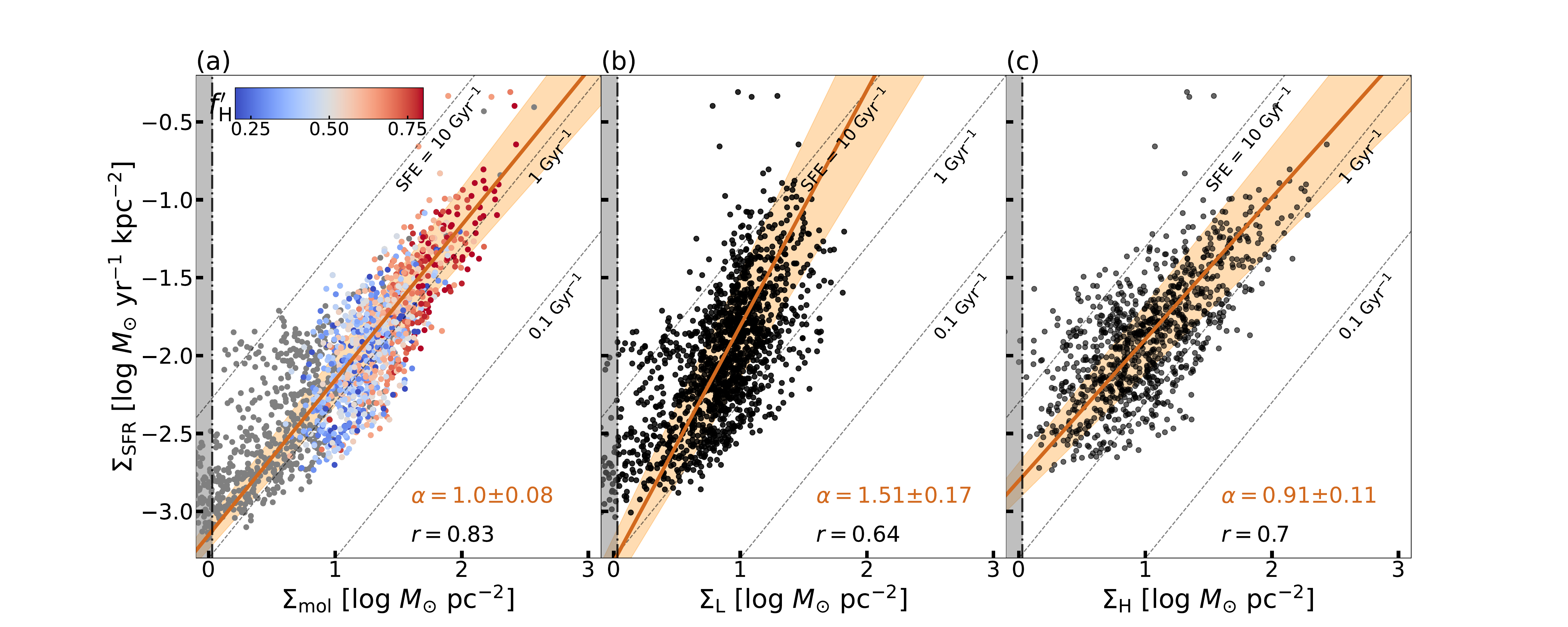}
    \caption{Relationship between the $\Sigma_{\rm SFR}$ and molecular gas surface density at a spatial scale of 550 pc. Panel (a) shows the $\Sigma_{\rm SFR}$ versus $\Sigma_{\rm mol}$ relation. The color scale represents the fraction of the high-density component $f_{\rm H}^{\prime}$. Gray points indicate all data points, including those with only an L-LN component. Panel (b) shows the $\Sigma_{\rm SFR}$ versus $\Sigma_{\rm L}$ relation. Panel (c) shows the $\Sigma_{\rm SFR}$ versus $\Sigma_{\rm H}$ relation. Diagonal dotted lines indicate constant SFEs, corresponding to gas consumption rates of 0.1$\%$, 1$\%$, and 10$\%$ within $10^7$ years. The vertical dot-dashed line with a gray shaded zone marks the $1\sigma$ sensitivity limit. The orange line with shading shows the best-fit relation with the associated uncertainties, derived using orthogonal distance regression. The slope ($\alpha$) and correlation coefficient ($r$) are shown in the lower right corner of each panel.}
    \label{fig:KSlaw}
\end{figure*}

The L-LN and H-LN component surface densities are defined as
\begin{equation}
\begin{split}
&\Sigma_{\rm L}\ [M_{\odot}\ {\rm pc}^{-2}]  =\frac{M_{\rm L}}{A_{\rm GDH}/\cos i}\ ,\\
&\Sigma_{\rm H}\ [M_{\odot}\ {\rm pc}^{-2}]  =\frac{M_{\rm H}}{A_{\rm GDH}/\cos i}\ ,
\end{split}
\end{equation}
where $M_{\rm L}$ and $M_{\rm H}$ are masses derived from the face-on surface density $\Sigma_{\rm mol}$ and the denominators are the deprojected GDH-cell area. We measured $\Sigma_{\rm SFR}$ from a combination of FUV and 24~$\micron$ maps. The FUV traces recent unobscured star formation, while the 24~$\micron$ emission captures the dust-obscured component by re-radiation from heated dust. This approach, validated for both individual star-forming regions and large areas of galactic disks \citep{CalzettiETAL2007,BigielETAL2008}, mitigates the bias in SFR estimates caused by dust attenuation in the FUV. We adopted a similar approach and combined FUV and 24~$\micron$ using the following formula:
\begin{equation}
\begin{split}
    \centering
    &\Sigma_{\rm SFR} = \Sigma_{\rm SFR}^{\rm FUV} + \Sigma_{\rm SFR}^{24\micron},\\
    &\Sigma_{\rm SFR}^{\rm FUV} = 8.1\times10^{-2}\ I_{\rm FUV}\cos i,\\
    &\Sigma_{\rm SFR}^{24\micron} = 3.2\times10^{-3}\ I_{24\micron}\cos i,
\end{split}
\end{equation}
where $\Sigma_{\rm SFR}^{\rm FUV}$ and $\Sigma_{\rm SFR}^{24\micron}$ are in $M_{\odot}\,{\rm yr^{-1}\,kpc^{-2}}$, and $I_{\rm FUV}$ and $I_{24\micron}$ are the GDH cell-averaged GALEX FUV and Spitzer 24~$\micron$ intensities in MJy\,sr$^{-1}$, respectively. Fig.~\ref{fig:SFR-map} shows the spatial distribution of the derived $\Sigma_{\rm SFR}$. Its overall spatial pattern is similar to that of $f^{\prime}_{\rm H}$ (see Fig.~\ref{fig:spatial_dist}c), suggesting that star formation is preferentially enhanced in regions with a high fraction of the H-LN component.

Fig.~\ref{fig:KSlaw} shows the KS relations between $\Sigma_{\rm SFR}$ and each component of molecular gas surface density at a spatial scale of 550~pc. Lines of constant star formation efficiency (${\rm SFE}\equiv\Sigma_{\rm SFR}/\Sigma_{\rm mol}\ [{\rm yr^{-1}}]$) are overlaid for reference. In this analysis, the slope and correlation coefficient are calculated using only the data above the sensitivity limits adopted in the GDH analysis. 

In Fig.~\ref{fig:KSlaw}, panel (a) shows $\Sigma_{\rm SFR}$ plotted against the molecular-gas surface density averaged over each GDH cell, with the colour scale representing $f_{\rm H}^{\prime}$. This GDH resolution is larger than the boundary spatial resolutions at which the KS relation breaks \citep[80–250 pc;][]{OnoderaETAL2010}. A previous sub-kpc scale study of M83 \citep{Morokuma-MatsuiETAL2017} reported a super-linear molecular KS slope of $\alpha\sim1.45$, with a typical sensitivity around $\Sigma_{\rm mol}\sim25~[M_\odot~{\rm pc^{-2}}]$ and a field of view limited to the inner disk. By contrast, our ALMA-based analysis reaches $\Sigma_{\rm mol}\sim1.25~[M_\odot~{\rm pc^{-2}}]$ more than an order of magnitude deeper and yields an overall slope that is nearly linear, $\alpha = 1.0 \pm 0.08$ (with $r=0.83$). We attribute this difference primarily to the improved sensitivity and wider FoV of our data, which include much lower $\Sigma_{\rm mol}$ and extend into the outer disk.

\subsection{Multi-component molecular gas and star formation relation}
The KS relation can be used as a tool to probe the multi-phase structure of the ISM by connecting changes in ISM properties with star formation activity \citep[e.g.,][]{BigielETAL2008,Morokuma-MatsuiETAL2017}. We therefore constructed KS relations for the total molecular gas and for each GDH component (Fig.~\ref{fig:KSlaw}). Panel (a) shows the relation for the total molecular gas—the conventional molecular KS relation, which exhibits a tight, nearly linear trend with a slope of $\alpha = 1.00 \pm 0.08$, consistent with classical results. The L-LN components, however, exhibit distinct behaviors. Panel (b) shows the relation for $\Sigma_{\rm L}$, which has a steeper slope of $\alpha = 1.51 \pm 0.17$ and a weaker correlation ($r = 0.64$), which may indicate that the L-LN is not tightly coupled to ongoing star formation at this scale. In contrast, panel (c) shows the relation for $\Sigma_{\rm H}$ which is nearly linear with a slope of $\alpha = 0.91 \pm 0.11$ and a slightly higher correlation ($r = 0.70$) compared to the L-LN component, a pattern that could be consistent with a comparatively closer connection between the H-LN phase and star formation activity. 

To interpret these relations, it is important to note that the L-LN/H-LN components identified in this work should not be interpreted simply as distinct thermodynamical density phases. At the present $\sim$40 pc resolution, they more directly describe how the molecular-gas surface-density distribution is organised within each 550 pc GDH cell. In this sense, the L-LN and H-LN components represent lower- and higher-surface-density structures, and their relative contributions may reflect not only intrinsic gas density but also the filling factor of molecular structures within the beam. The slope of the L-LN component is close to those found for \HI\ gas \citep[e.g.,][]{BigielETAL2008}. The L-LN component appears to saturate at a surface density of $\sim 30~M_\odot~\mathrm{pc}^{-2}$ (Fig.~\ref{fig:KSlaw}b), a value comparable in order of magnitude to the canonical \HI\ saturation threshold reported for nearby galaxies and the Milky Way \citep[$\sim$10--60~$M_\odot~\mathrm{pc}^{-2}$;][]{BigielETAL2008,ParkETAL2023}. This correspondence suggests that the L-LN component traces a more spatially extended molecular reservoir that is only loosely coupled to ongoing star formation. By contrast, the H-LN component shows a tighter and nearly linear relation with $\Sigma_{\rm SFR}$, indicating that star formation is more closely associated with regions where high-surface-density molecular structures make a larger contribution within a GDH cell.

In combination with the spatial distributions of $f'_{\rm H}$ and $\Sigma_{\rm SFR}$, these results suggest a coherent environmental picture. Regions of elevated $\Sigma_{\rm SFR}$ preferentially coincide with regions of high $f^{\prime}_{\rm H}$, especially along the spiral arms of M83 (see Fig.~\ref{fig:spatial_dist}c and Fig.~\ref{fig:SFR-map}). The H-LN component therefore appears to trace molecular structures more directly linked to star formation, whereas the L-LN component represents a more extended reservoir distributed over the disk.

\section{Conclusions}
\label{sec:conclusions}
We present a sub-kpc scale statistical analysis of the molecular-gas surface-density structure in the barred spiral galaxy M83 using the gas density histogram (GDH) method. Specifically, we tile the disk with GDH cells of $550~\mathrm{pc}\times550~\mathrm{pc}\times100~\mathrm{km~s^{-1}}$. For every GDH cell we convert the CO brightness to $\Sigma_{\rm mol}$ and then construct the histogram of $\log\Sigma_{\rm mol}$. The histograms are fitted as one or two LN components, representing lower and higher-surface-density molecular components, referred to as L-LN and H-LN, respectively. We find that about 40$\%$ of the total molecular-gas mass is contained in the L-LN component. Our main conclusions are as follows:

\begin{enumerate}
    \item The GDH method has proven to be a powerful tool for characterising the surface-density distribution of molecular gas. In the case of M83, the GDHs in almost all regions were well approximated by one or two LN components.
    
    \item The spatial distribution of the H-LN component mass ($M_{\rm H}$) is significantly more structured than that of the L-LN component ($M_{\rm L}$). This structure is well correlating with spiral arms. Furthermore, the radial profiles show that $M_{\rm H}$ is more prominent than $M_{\rm L}$ in the inner disk, while its relative contribution decreases toward the outer disk. This trend indicates that the relative contribution of the H-LN component decreases with radius, such that the L-LN component becomes increasingly important in the outer disk.
    
    \item The mass fraction of the H-LN component ($f^{\prime}_{\rm H}$) shows coherent structures across the disk, peaking along prominent spiral structures. This suggests that spiral-arm environments are associated with an increased contribution of the H-LN component relative to the L-LN component. This result is consistent with our previous study of the Milky Way \citep{MatsusakaETAL2024}, which also revealed enhanced area-based H-LN fractions along spiral features identified on the $l$–$v$ plane, despite the galaxy's edge-on orientation.
    
    \item $\Sigma_{\rm SFR}$ correlates more strongly with the H-LN component than with the L-LN component: the H-LN KS relation is nearly linear, comparable to the conventional molecular KS relation's trend, whereas the L-LN relation is steeper and shows a surface-density saturation at $\sim30~M_\odot\,\mathrm{pc^{-2}}$, reminiscent of the behaviour commonly reported for \HI. This suggests that star formation is more closely linked to regions where the H-LN component makes a larger contribution within a GDH cell, whereas the L-LN component represents a more spatially extended molecular reservoir only loosely coupled to ongoing star formation.
\end{enumerate}

These results provide direct observational evidence that the molecular ISM in M83 consists of multiple molecular gas components with distinct spatial distributions and different levels of association with star formation. Because the GDH method is independent of cloud identification, it offers a powerful and relatively unbiased diagnostic of molecular-gas structure. Future studies should aim to link the GDH-derived components more directly to the underlying physical conditions by jointly examining CO($J{=}2$--1)/CO($J{=}1$--0) ratios, dense-gas fractions traced by high-critical-density molecular lines, and GDH parameters. Moreover, applying the method with smaller GDH cells (e.g., $\lesssim100$~pc), enabled by future observations at a resolution of a few parsecs, will provide deeper insight into how the structure of molecular gas varies across galactic environments.

\section*{Acknowledgements}
We thank the anonymous referee for a careful reading of the manuscript and for constructive comments that helped improve the paper. We are deeply grateful to Dr. Jin Koda for generously providing the observational data that formed the basis of this work. We are grateful to Drs. Kotaro Kohno and Yuri Nishimura for the useful discussions. F.M. is supported by JSPS KAKENHI grant Nos. JP23K13142 and JP23K20035. This paper makes use of the following ALMA data: ADS / JAO.ALMA $\#$2017.1.00079.S. ALMA is a partnership of ESO (representing its member states), NSF (USA) and NINS (Japan), together with NRC (Canada), MOST and ASIAA (Taiwan), and KASI (Republic of Korea), in cooperation with the Republic of Chile. The Joint ALMA Observatory is operated by ESO, AUI / NRAO and NAOJ. Data analysis was in part carried out on the Multi-wavelength Data Analysis System operated by the Astronomy Data Center (ADC), National Astronomical Observatory of Japan. 
We acknowledge the use of NASA’s SkyView facility
\url{http://skyview.gsfc.nasa.gov} located at NASA Goddard Space Flight Center. This research used \textsc{astropy},\footnote{\url{http://www.astropy.org}} a community-developed core Python package for Astronomy \citep{astropy_2013,astropy_2018}, \textsc{matplotlib}, a Python package for visualisation \citep{Hunter2007}, \textsc{numpy}, a Python package for scientific computing \citep{HarrisETAL2020}, and Overleaf, a collaborative tool.

\section*{Data Availability}
The primary $^{12}$CO($J$=  1--0) dataset for M83 used in this study is available from the ALMA Science Archive under the project identifier \texttt{ADS/JAO.ALMA\#2017.1.00079.S}. The observations and data reduction are described by \citet{KodaETAL2023} and \cite{KodaETAL2025}. Publicly archived ancillary data were obtained from: DustPedia \citep{ClarkETAL2018}, \url{http://dustpedia.astro.noa.gr/}; THINGS \citep{WalterETAL2008}, \url{https://www2.mpia-hd.mpg.de/THINGS/Overview.html}; and NASA SkyView \citep{McGlynnETAL1998}, \url{https://skyview.gsfc.nasa.gov/}. The Milky Way results presented in \citet{MatsusakaETAL2024} use data from the FUGIN project, available at \url{https://nro-fugin.github.io/}. Derived data products generated in this study (e.g., GDH maps, masks, and region catalogs) will be shared on reasonable request to the corresponding author.



\bibliographystyle{mnras}
\bibliography{reference} 

@ARTICLE{KritsukETAL2011,
       author = {{Kritsuk}, Alexei G. and {Norman}, Michael L. and {Wagner}, Rick},
        title = "{On the Density Distribution in Star-forming Interstellar Clouds}",
      journal = {\apjl},
         year = 2011,
        month = jan,
       volume = {727},
       number = {1},
          eid = {L20},
        pages = {L20},
          doi = {10.1088/2041-8205/727/1/L20},
archivePrefix = {arXiv},
       eprint = {1007.2950},
 primaryClass = {astro-ph.GA},
       adsurl = {https://ui.adsabs.harvard.edu/abs/2011ApJ...727L..20K}
}

@ARTICLE{FederrathETAL2008,
       author = {{Federrath}, Christoph and {Klessen}, Ralf S. and {Schmidt}, Wolfram},
        title = "{The Density Probability Distribution in Compressible Isothermal Turbulence: Solenoidal versus Compressive Forcing}",
      journal = {\apjl},
         year = 2008,
        month = dec,
       volume = {688},
       number = {2},
        pages = {L79},
          doi = {10.1086/595280},
archivePrefix = {arXiv},
       eprint = {0808.0605},
 primaryClass = {astro-ph},
       adsurl = {https://ui.adsabs.harvard.edu/abs/2008ApJ...688L..79F}
}

@ARTICLE{PadoanETAL1997,
       author = {{Padoan}, Paolo and {Jones}, Bernard J.~T. and {Nordlund}, {\r{A}}ke P.},
        title = "{Supersonic Turbulence in the Interstellar Medium: Stellar Extinction Determinations as Probes of the Structure and Dynamics of Dark Clouds}",
      journal = {\apj},
         year = 1997,
        month = jan,
       volume = {474},
       number = {2},
        pages = {730-734},
          doi = {10.1086/303482},
archivePrefix = {arXiv},
       eprint = {astro-ph/9603061},
 primaryClass = {astro-ph},
       adsurl = {https://ui.adsabs.harvard.edu/abs/1997ApJ...474..730P}
}

@ARTICLE{TremblinETL2014,
       author = {{Tremblin}, P. and {Schneider}, N. and {Minier}, V. and {Didelon}, P. and {Hill}, T. and {Anderson}, L.~D. and {Motte}, F. and {Zavagno}, A. and {Andr{\'e}}, Ph. and {Arzoumanian}, D. and {Audit}, E. and {Benedettini}, M. and {Bontemps}, S. and {Csengeri}, T. and {Di Francesco}, J. and {Giannini}, T. and {Hennemann}, M. and {Nguyen Luong}, Q. and {Marston}, A.~P. and {Peretto}, N. and {Rivera-Ingraham}, A. and {Russeil}, D. and {Rygl}, K.~L.~J. and {Spinoglio}, L. and {White}, G.~J.},
        title = "{Ionization compression impact on dense gas distribution and star formation. Probability density functions around H II regions as seen by Herschel}",
      journal = {\aap},
         year = 2014,
        month = apr,
       volume = {564},
          eid = {A106},
        pages = {A106},
          doi = {10.1051/0004-6361/201322700},
archivePrefix = {arXiv},
       eprint = {1401.7333},
 primaryClass = {astro-ph.GA},
       adsurl = {https://ui.adsabs.harvard.edu/abs/2014A&A...564A.106T}
}

@ARTICLE{SchneiderETAL2025,
       author = {{Schneider}, Nicola and {Ossenkopf-Okada}, Volker and {R{\"o}llig}, Markus and {Seifried}, Daniel and {Klessen}, Ralf S. and {Kritsuk}, Alexei G. and {Keilmann}, Eduard and {Dannhauer}, Simon and {Bonne}, Lars and {Glover}, Simon C.~O.},
        title = "{The HI-to-H$_{2}$ transition in the Draco cloud}",
      journal = {\aap},
         year = 2025,
        month = jul,
       volume = {699},
          eid = {A354},
        pages = {A354},
          doi = {10.1051/0004-6361/202555308},
archivePrefix = {arXiv},
       eprint = {2507.06131},
 primaryClass = {astro-ph.GA},
       adsurl = {https://ui.adsabs.harvard.edu/abs/2025A&A...699A.354S}
}

@ARTICLE{Baba2015,
       author = {{Baba}, Junichi},
        title = "{Short-term dynamical evolution of grand-design spirals in barred galaxies}",
      journal = {\mnras},
         year = 2015,
        month = dec,
       volume = {454},
       number = {3},
        pages = {2954-2964},
          doi = {10.1093/mnras/stv2220},
archivePrefix = {arXiv},
       eprint = {1509.07239},
 primaryClass = {astro-ph.GA},
       adsurl = {https://ui.adsabs.harvard.edu/abs/2015MNRAS.454.2954B}
}

@ARTICLE{QuerejetaETAL2025,
       author = {{Querejeta}, Miguel and {Meidt}, Sharon E. and {Cao}, Yixian and {Colombo}, Dario and {Emsellem}, Eric and {Garc{\'\i}a-Burillo}, Santiago and {Klessen}, Ralf S. and {Koch}, Eric W. and {Leroy}, Adam K. and {Ruiz-Garc{\'\i}a}, Marina and {Schinnerer}, Eva and {Smith}, Rowan and {Stuber}, Sophia and {Thorp}, Mallory and {Williams}, Thomas G. and {Boquien}, M{\'e}d{\'e}ric and {Dale}, Daniel A. and {Faesi}, Chris and {Gleis}, Damian R. and {Grasha}, Kathryn and {Hughes}, Annie and {Jim{\'e}nez-Donaire}, Mar{\'\i}a J. and {Kreckel}, Kathryn and {Liu}, Daizhong and {Neumann}, Justus and {Pan}, Hsi-An and {Pinna}, Francesca and {Razza}, Alessandro and {Saito}, Toshiki and {Sun}, Jiayi and {Usero}, Antonio},
        title = "{Azimuthal offsets in spiral arms of nearby galaxies}",
      journal = {\aap},
         year = 2025,
        month = sep,
       volume = {701},
          eid = {A183},
        pages = {A183},
          doi = {10.1051/0004-6361/202556175},
archivePrefix = {arXiv},
       eprint = {2509.01668},
 primaryClass = {astro-ph.GA},
       adsurl = {https://ui.adsabs.harvard.edu/abs/2025A&A...701A.183Q}
}

@ARTICLE{NiETAL2025,
       author = {{Ni}, Yang and {Li}, Hui and {Vogelsberger}, Mark and {Sales}, Laura V. and {Marinacci}, Federico and {Torrey}, Paul},
        title = "{The life cycle of giant molecular clouds in simulated Milky Way-mass galaxies}",
      journal = {\aap},
         year = 2025,
        month = jul,
       volume = {699},
          eid = {A282},
        pages = {A282},
          doi = {10.1051/0004-6361/202554126},
archivePrefix = {arXiv},
       eprint = {2502.12256},
 primaryClass = {astro-ph.GA},
       adsurl = {https://ui.adsabs.harvard.edu/abs/2025A&A...699A.282N}
}

@ARTICLE{ColmanETAL2024,
       author = {{Colman}, Tine and {Brucy}, No{\'e} and {Girichidis}, Philipp and {Glover}, Simon C.~O. and {Benedettini}, Milena and {Soler}, Juan D. and {Tress}, Robin G. and {Traficante}, Alessio and {Hennebelle}, Patrick and {Klessen}, Ralf S. and {Molinari}, Sergio and {Miville-Desch{\^e}nes}, Marc-Antoine},
        title = "{Cloud properties across spatial scales in simulations of the interstellar medium}",
      journal = {\aap},
         year = 2024,
        month = jun,
       volume = {686},
          eid = {A155},
        pages = {A155},
          doi = {10.1051/0004-6361/202348983},
archivePrefix = {arXiv},
       eprint = {2403.00512},
 primaryClass = {astro-ph.GA},
       adsurl = {https://ui.adsabs.harvard.edu/abs/2024A&A...686A.155C}
}

@ARTICLE{EgusaETAL2009,
       author = {{Egusa}, Fumi and {Kohno}, Kotaro and {Sofue}, Yoshiaki and {Nakanishi}, Hiroyuki and {Komugi}, Shinya},
        title = "{Determining Star Formation Timescale and Pattern Speed in Nearby Spiral Galaxies}",
      journal = {\apj},
         year = 2009,
        month = jun,
       volume = {697},
       number = {2},
        pages = {1870-1891},
          doi = {10.1088/0004-637X/697/2/1870},
archivePrefix = {arXiv},
       eprint = {0904.3121},
 primaryClass = {astro-ph.CO},
       adsurl = {https://ui.adsabs.harvard.edu/abs/2009ApJ...697.1870E}
}

@ARTICLE{EgusaETAL2004,
       author = {{Egusa}, Fumi and {Sofue}, Yoshiaki and {Nakanishi}, Hiroyuki},
        title = "{Offsets between H{\ensuremath{\alpha}} and CO Arms of a Spiral Galaxy, NGC 4254: A New Method for Determining the Pattern Speed of Spiral Galaxies}",
      journal = {\pasj},
         year = 2004,
        month = dec,
       volume = {56},
        pages = {L45-L48},
          doi = {10.1093/pasj/56.6.L45},
archivePrefix = {arXiv},
       eprint = {astro-ph/0410469},
 primaryClass = {astro-ph},
       adsurl = {https://ui.adsabs.harvard.edu/abs/2004PASJ...56L..45E}
}

@ARTICLE{HirotaETAL2024,
       author = {{Hirota}, Akihiko and {Koda}, Jin and {Egusa}, Fumi and {Sawada}, Tsuyoshi and {Sakamoto}, Kazushi and {Heyer}, Mark and {Lee}, Amanda M. and {Maeda}, Fumiya and {Boissier}, Samuel and {Calzetti}, Daniela and {Elmegreen}, Bruce G. and {Harada}, Nanase and {Ho}, Luis C. and {Kobayashi}, Masato I.~N. and {Kuno}, Nario and {Madore}, Barry F. and {Mart{\'\i}n}, Sergio and {Donovan Meyer}, Jennifer and {Muraoka}, Kazuyuki and {Watanabe}, Yoshimasa},
        title = "{Whole-disk Sampling of Molecular Clouds in M83}",
      journal = {\apj},
         year = 2024,
        month = dec,
       volume = {976},
       number = {2},
          eid = {198},
        pages = {198},
          doi = {10.3847/1538-4357/ad8228},
archivePrefix = {arXiv},
       eprint = {2410.05424},
 primaryClass = {astro-ph.GA},
       adsurl = {https://ui.adsabs.harvard.edu/abs/2024ApJ...976..198H}
}

@ARTICLE{KonishiETAL2024,
       author = {{Konishi}, Ayu and {Muraoka}, Kazuyuki and {Tokuda}, Kazuki and {Fujita}, Shinji and {Fukui}, Yasuo and {Yamada}, Rin I. and {Demachi}, Fumika and {Tachihara}, Kengo and {Kobayashi}, Masato I.~N. and {Kuno}, Nario and {Tsuge}, Kisetsu and {Sano}, Hidetoshi and {Miura}, Rie E. and {Kawamura}, Akiko and {Onishi}, Toshikazu},
        title = "{ACA CO(J = 2-1) mapping of the nearest spiral galaxy M 33. II. Exploring the evolution of giant molecular clouds}",
      journal = {\pasj},
         year = 2024,
        month = oct,
       volume = {76},
       number = {5},
        pages = {1098-1121},
          doi = {10.1093/pasj/psae073},
archivePrefix = {arXiv},
       eprint = {2407.17018},
 primaryClass = {astro-ph.GA},
       adsurl = {https://ui.adsabs.harvard.edu/abs/2024PASJ...76.1098K}
}

@ARTICLE{CASAteam2022,
       author = {{CASA Team} and {Bean}, Ben and {Bhatnagar}, Sanjay and {Castro}, Sandra and {Donovan Meyer}, Jennifer and {Emonts}, Bjorn and {Garcia}, Enrique and {Garwood}, Robert and {Golap}, Kumar and {Gonzalez Villalba}, Justo and {Harris}, Pamela and {Hayashi}, Yohei and {Hoskins}, Josh and {Hsieh}, Mingyu and {Jagannathan}, Preshanth and {Kawasaki}, Wataru and {Keimpema}, Aard and {Kettenis}, Mark and {Lopez}, Jorge and {Marvil}, Joshua and {Masters}, Joseph and {McNichols}, Andrew and {Mehringer}, David and {Miel}, Renaud and {Moellenbrock}, George and {Montesino}, Federico and {Nakazato}, Takeshi and {Ott}, Juergen and {Petry}, Dirk and {Pokorny}, Martin and {Raba}, Ryan and {Rau}, Urvashi and {Schiebel}, Darrell and {Schweighart}, Neal and {Sekhar}, Srikrishna and {Shimada}, Kazuhiko and {Small}, Des and {Steeb}, Jan-Willem and {Sugimoto}, Kanako and {Suoranta}, Ville and {Tsutsumi}, Takahiro and {van Bemmel}, Ilse M. and {Verkouter}, Marjolein and {Wells}, Akeem and {Xiong}, Wei and {Szomoru}, Arpad and {Griffith}, Morgan and {Glendenning}, Brian and {Kern}, Jeff},
        title = "{CASA, the Common Astronomy Software Applications for Radio Astronomy}",
      journal = {\pasp},
         year = 2022,
        month = nov,
       volume = {134},
       number = {1041},
          eid = {114501},
        pages = {114501},
          doi = {10.1088/1538-3873/ac9642},
archivePrefix = {arXiv},
       eprint = {2210.02276},
 primaryClass = {astro-ph.IM},
       adsurl = {https://ui.adsabs.harvard.edu/abs/2022PASP..134k4501C}
}

@ARTICLE{OnoderaETAL2010,
       author = {{Onodera}, Sachiko and {Kuno}, Nario and {Tosaki}, Tomoka and {Kohno}, Kotaro and {Nakanishi}, Kouichiro and {Sawada}, Tsuyoshi and {Muraoka}, Kazuyuki and {Komugi}, Shinya and {Miura}, Rie and {Kaneko}, Hiroyuki and {Hirota}, Akihiko and {Kawabe}, Ryohei},
        title = "{Breakdown of Kennicutt-Schmidt Law at Giant Molecular Cloud Scales in M33}",
      journal = {\apjl},
         year = 2010,
        month = oct,
       volume = {722},
       number = {2},
        pages = {L127-L131},
          doi = {10.1088/2041-8205/722/2/L127},
archivePrefix = {arXiv},
       eprint = {1009.1971},
 primaryClass = {astro-ph.GA},
       adsurl = {https://ui.adsabs.harvard.edu/abs/2010ApJ...722L.127O}
}

@ARTICLE{Morokuma-MatsuiETAL2017,
       author = {{Morokuma-Matsui}, Kana and {Muraoka}, Kazuyuki},
        title = "{Kennicutt-Schmidt Relation Variety and Star-forming Cloud Fraction}",
      journal = {\apj},
         year = 2017,
        month = mar,
       volume = {837},
       number = {2},
          eid = {137},
        pages = {137},
          doi = {10.3847/1538-4357/aa6115},
archivePrefix = {arXiv},
       eprint = {1702.04820},
 primaryClass = {astro-ph.GA},
       adsurl = {https://ui.adsabs.harvard.edu/abs/2017ApJ...837..137M}
}

@ARTICLE{NeumannETAL2023,
       author = {{Neumann}, Lukas and {Gallagher}, Molly J. and {Bigiel}, Frank and {Leroy}, Adam K. and {Barnes}, Ashley T. and {Usero}, Antonio and {den Brok}, Jakob S. and {Belfiore}, Francesco and {Be{\v{s}}li{\'c}}, Ivana and {Cao}, Yixian and {Chevance}, M{\'e}lanie and {Dale}, Daniel A. and {Eibensteiner}, Cosima and {Glover}, Simon C.~O. and {Grasha}, Kathryn and {Henshaw}, Jonathan D. and {Jim{\'e}nez-Donaire}, Mar{\'\i}a J. and {Klessen}, Ralf S. and {Kruijssen}, J.~M. Diederik and {Liu}, Daizhong and {Meidt}, Sharon and {Pety}, J{\'e}r{\^o}me and {Puschnig}, Johannes and {Querejeta}, Miguel and {Rosolowsky}, Erik and {Schinnerer}, Eva and {Schruba}, Andreas and {Sormani}, Mattia C. and {Sun}, Jiayi and {Teng}, Yu-Hsuan and {Williams}, Thomas G.},
        title = "{The ALMOND survey: molecular cloud properties and gas density tracers across 25 nearby spiral galaxies with ALMA}",
      journal = {\mnras},
         year = 2023,
        month = may,
       volume = {521},
       number = {3},
        pages = {3348-3383},
          doi = {10.1093/mnras/stad424},
archivePrefix = {arXiv},
       eprint = {2302.03042},
 primaryClass = {astro-ph.GA},
       adsurl = {https://ui.adsabs.harvard.edu/abs/2023MNRAS.521.3348N}
}

@ARTICLE{KohnoETAL2024,
       author = {{Kohno}, Mikito and {Sofue}, Yoshiaki},
        title = "{The CO-to-H$_{2}$ conversion factor of Galactic giant molecular clouds using CO isotopologues: high-resolution X$_{CO}$ maps}",
      journal = {\mnras},
         year = 2024,
        month = jan,
       volume = {527},
       number = {3},
        pages = {9290-9302},
          doi = {10.1093/mnras/stad3648},
archivePrefix = {arXiv},
       eprint = {2311.13760},
 primaryClass = {astro-ph.GA},
       adsurl = {https://ui.adsabs.harvard.edu/abs/2024MNRAS.527.9290K}
}

@ARTICLE{LeeETAL2024,
       author = {{Lee}, Amanda M. and {Koda}, Jin and {Hirota}, Akihiko and {Egusa}, Fumi and {Heyer}, Mark},
        title = "{The CO-to-H$_{2}$ Conversion Factor in the Barred Spiral Galaxy M83}",
      journal = {\apj},
         year = 2024,
        month = jun,
       volume = {968},
       number = {2},
          eid = {97},
        pages = {97},
          doi = {10.3847/1538-4357/ad40a0},
archivePrefix = {arXiv},
       eprint = {2404.14503},
 primaryClass = {astro-ph.GA},
       adsurl = {https://ui.adsabs.harvard.edu/abs/2024ApJ...968...97L}
}

@ARTICLE{ClarkETAL2018,
       author = {{Clark}, C.~J.~R. and {Verstocken}, S. and {Bianchi}, S. and {Fritz}, J. and {Viaene}, S. and {Smith}, M.~W.~L. and {Baes}, M. and {Casasola}, V. and {Cassara}, L.~P. and {Davies}, J.~I. and {De Looze}, I. and {De Vis}, P. and {Evans}, R. and {Galametz}, M. and {Jones}, A.~P. and {Lianou}, S. and {Madden}, S. and {Mosenkov}, A.~V. and {Xilouris}, M.},
        title = "{DustPedia: Multiwavelength photometry and imagery of 875 nearby galaxies in 42 ultraviolet-microwave bands}",
      journal = {\aap},
         year = 2018,
        month = jan,
       volume = {609},
          eid = {A37},
        pages = {A37},
          doi = {10.1051/0004-6361/201731419},
archivePrefix = {arXiv},
       eprint = {1708.05335},
 primaryClass = {astro-ph.GA},
       adsurl = {https://ui.adsabs.harvard.edu/abs/2018A&A...609A..37C}
}

@ARTICLE{JimETAL2019,
       author = {{Jim{\'e}nez-Donaire}, Mar{\'\i}a J. and {Bigiel}, F. and {Leroy}, A.~K. and {Usero}, A. and {Cormier}, D. and {Puschnig}, J. and {Gallagher}, M. and {Kepley}, A. and {Bolatto}, A. and {Garc{\'\i}a-Burillo}, S. and {Hughes}, A. and {Kramer}, C. and {Pety}, J. and {Schinnerer}, E. and {Schruba}, A. and {Schuster}, K. and {Walter}, F.},
        title = "{EMPIRE: The IRAM 30 m Dense Gas Survey of Nearby Galaxies}",
      journal = {\apj},
         year = 2019,
        month = aug,
       volume = {880},
       number = {2},
          eid = {127},
        pages = {127},
          doi = {10.3847/1538-4357/ab2b95},
archivePrefix = {arXiv},
       eprint = {1906.08779},
 primaryClass = {astro-ph.GA},
       adsurl = {https://ui.adsabs.harvard.edu/abs/2019ApJ...880..127J}
}

@ARTICLE{KodaETAL2025,
       author = {{Koda}, Jin and {Egusa}, Fumi and {Hirota}, Akihiko and {Lee}, Amanda M. and {Sawada}, Tsuyoshi and {Maeda}, Fumiya},
        title = "{Dynamically Driven Evolution of Molecular Gas in the Barred Spiral Galaxy M83 Traced by CO J = 2{\textendash}1/1{\textendash}0 Line Ratio Variations}",
      journal = {\apj},
         year = 2025,
        month = jun,
       volume = {986},
       number = {1},
          eid = {29},
        pages = {29},
          doi = {10.3847/1538-4357/add1dc},
       adsurl = {https://ui.adsabs.harvard.edu/abs/2025ApJ...986...29K}
}

@ARTICLE{ShettyETAL2014,
       author = {{Shetty}, Rahul and {Clark}, Paul C. and {Klessen}, Ralf S.},
        title = "{Interpreting the sub-linear Kennicutt-Schmidt relationship: the case for diffuse molecular gas}",
      journal = {\mnras},
         year = 2014,
        month = aug,
       volume = {442},
       number = {3},
        pages = {2208-2215},
          doi = {10.1093/mnras/stu919},
archivePrefix = {arXiv},
       eprint = {1404.5964},
 primaryClass = {astro-ph.GA},
       adsurl = {https://ui.adsabs.harvard.edu/abs/2014MNRAS.442.2208S}
}

@ARTICLE{ParkETAL2023,
       author = {{Park}, Geumsook and {Koo}, Bon-Chul and {Kim}, Kee-Tae and {Elmegreen}, Bruce},
        title = "{Neutral Atomic and Molecular Clouds and Star Formation in the Outer Carina Arm}",
      journal = {\apj},
         year = 2023,
        month = sep,
       volume = {955},
       number = {1},
          eid = {59},
        pages = {59},
          doi = {10.3847/1538-4357/acebda},
archivePrefix = {arXiv},
       eprint = {2308.01577},
 primaryClass = {astro-ph.GA},
       adsurl = {https://ui.adsabs.harvard.edu/abs/2023ApJ...955...59P}
}

@INPROCEEDINGS{McGlynnETAL1998,
       author = {{McGlynn}, T. and {Scollick}, K. and {White}, N.},
        title = "{SKYVIEW:The Multi-Wavelength Sky on the Internet}",
    booktitle = {New Horizons from Multi-Wavelength Sky Surveys},
         year = 1998,
       editor = {{McLean}, Brian J. and {Golombek}, Daniel A. and {Hayes}, Jeffrey J.~E. and {Payne}, Harry E.},
       series = {IAU Symposium},
       volume = {179},
        month = jan,
        pages = {465},
       adsurl = {https://ui.adsabs.harvard.edu/abs/1998IAUS..179..465M}
}

@ARTICLE{KhullarETAL2021,
       author = {{Khullar}, Shivan and {Federrath}, Christoph and {Krumholz}, Mark R. and {Matzner}, Christopher D.},
        title = "{The density structure of supersonic self-gravitating turbulence}",
      journal = {\mnras},
         year = 2021,
        month = nov,
       volume = {507},
       number = {3},
        pages = {4335-4351},
          doi = {10.1093/mnras/stab1914},
archivePrefix = {arXiv},
       eprint = {2107.00725},
 primaryClass = {astro-ph.GA},
       adsurl = {https://ui.adsabs.harvard.edu/abs/2021MNRAS.507.4335K}
}

@ARTICLE{KodaETAL2009,
       author = {{Koda}, Jin and {Scoville}, Nick and {Sawada}, Tsuyoshi and {La Vigne}, Misty A. and {Vogel}, Stuart N. and {Potts}, Ashley E. and {Carpenter}, John M. and {Corder}, Stuartt A. and {Wright}, Melvyn C.~H. and {White}, Stephen M. and {Zauderer}, B. Ashley and {Patience}, Jenny and {Sargent}, Anneila I. and {Bock}, Douglas C.~J. and {Hawkins}, David and {Hodges}, Mark and {Kemball}, Athol and {Lamb}, James W. and {Plambeck}, Richard L. and {Pound}, Marc W. and {Scott}, Stephen L. and {Teuben}, Peter and {Woody}, David P.},
        title = "{Dynamically Driven Evolution of the Interstellar Medium in M51}",
      journal = {\apjl},
         year = 2009,
        month = aug,
       volume = {700},
       number = {2},
        pages = {L132-L136},
          doi = {10.1088/0004-637X/700/2/L132},
archivePrefix = {arXiv},
       eprint = {0907.1656},
 primaryClass = {astro-ph.CO},
       adsurl = {https://ui.adsabs.harvard.edu/abs/2009ApJ...700L.132K}
}

@ARTICLE{EgusaETAL2011,
       author = {{Egusa}, Fumi and {Koda}, Jin and {Scoville}, Nick},
        title = "{Molecular Gas Evolution Across a Spiral Arm in M51}",
      journal = {\apj},
         year = 2011,
        month = jan,
       volume = {726},
       number = {2},
          eid = {85},
        pages = {85},
          doi = {10.1088/0004-637X/726/2/85},
archivePrefix = {arXiv},
       eprint = {1011.3889},
 primaryClass = {astro-ph.CO},
       adsurl = {https://ui.adsabs.harvard.edu/abs/2011ApJ...726...85E}
}

@ARTICLE{Schmidt1959,
       author = {{Schmidt}, Maarten},
        title = "{The Rate of Star Formation.}",
      journal = {\apj},
         year = 1959,
        month = mar,
       volume = {129},
        pages = {243},
          doi = {10.1086/146614},
       adsurl = {https://ui.adsabs.harvard.edu/abs/1959ApJ...129..243S}
}

@ARTICLE{Kennicutt1998,
       author = {{Kennicutt}, Robert C., Jr.},
        title = "{The Global Schmidt Law in Star-forming Galaxies}",
      journal = {\apj},
         year = 1998,
        month = may,
       volume = {498},
       number = {2},
        pages = {541-552},
          doi = {10.1086/305588},
archivePrefix = {arXiv},
       eprint = {astro-ph/9712213},
 primaryClass = {astro-ph},
       adsurl = {https://ui.adsabs.harvard.edu/abs/1998ApJ...498..541K}
}

@ARTICLE{WalterETAL2008,
       author = {{Walter}, Fabian and {Brinks}, Elias and {de Blok}, W.~J.~G. and {Bigiel}, Frank and {Kennicutt}, Robert C., Jr. and {Thornley}, Michele D. and {Leroy}, Adam},
        title = "{THINGS: The H I Nearby Galaxy Survey}",
      journal = {\aj},
         year = 2008,
        month = dec,
       volume = {136},
       number = {6},
        pages = {2563-2647},
          doi = {10.1088/0004-6256/136/6/2563},
archivePrefix = {arXiv},
       eprint = {0810.2125},
 primaryClass = {astro-ph},
       adsurl = {https://ui.adsabs.harvard.edu/abs/2008AJ....136.2563W}
}

@ARTICLE{DameETAL2001,
       author = {{Dame}, T.~M. and {Hartmann}, Dap and {Thaddeus}, P.},
        title = "{The Milky Way in Molecular Clouds: A New Complete CO Survey}",
      journal = {\apj},
         year = 2001,
        month = feb,
       volume = {547},
       number = {2},
        pages = {792-813},
          doi = {10.1086/318388},
archivePrefix = {arXiv},
       eprint = {astro-ph/0009217},
 primaryClass = {astro-ph},
       adsurl = {https://ui.adsabs.harvard.edu/abs/2001ApJ...547..792D}
}

@ARTICLE{YanETAL2022,
       author = {{Yan}, Qing-Zeng and {Yang}, Ji and {Su}, Yang and {Sun}, Yan and {Zhou}, Xin and {Xu}, Ye and {Wang}, Hongchi and {Zhang}, Shaobo and {Chen}, Zhiwei},
        title = "{Dependence of Molecular Cloud Samples on Angular Resolution, Sensitivity, and Algorithms}",
      journal = {\aj},
         year = 2022,
        month = aug,
       volume = {164},
       number = {2},
          eid = {55},
        pages = {55},
          doi = {10.3847/1538-3881/ac77ea},
archivePrefix = {arXiv},
       eprint = {2206.05436},
 primaryClass = {astro-ph.GA},
       adsurl = {https://ui.adsabs.harvard.edu/abs/2022AJ....164...55Y}
}

@ARTICLE{LeroyETAL2021,
       author = {{Leroy}, Adam K. and {Schinnerer}, Eva and {Hughes}, Annie and {Rosolowsky}, Erik and {Pety}, J{\'e}r{\^o}me and {Schruba}, Andreas and {Usero}, Antonio and {Blanc}, Guillermo A. and {Chevance}, M{\'e}lanie and {Emsellem}, Eric and {Faesi}, Christopher M. and {Herrera}, Cinthya N. and {Liu}, Daizhong and {Meidt}, Sharon E. and {Querejeta}, Miguel and {Saito}, Toshiki and {Sandstrom}, Karin M. and {Sun}, Jiayi and {Williams}, Thomas G. and {Anand}, Gagandeep S. and {Barnes}, Ashley T. and {Behrens}, Erica A. and {Belfiore}, Francesco and {Benincasa}, Samantha M. and {Be{\v{s}}li{\'c}}, Ivana and {Bigiel}, Frank and {Bolatto}, Alberto D. and {den Brok}, Jakob S. and {Cao}, Yixian and {Chandar}, Rupali and {Chastenet}, J{\'e}r{\'e}my and {Chiang}, I-Da and {Congiu}, Enrico and {Dale}, Daniel A. and {Deger}, Sinan and {Eibensteiner}, Cosima and {Egorov}, Oleg V. and {Garc{\'\i}a-Rodr{\'\i}guez}, Axel and {Glover}, Simon C.~O. and {Grasha}, Kathryn and {Henshaw}, Jonathan D. and {Ho}, I. -Ting and {Kepley}, Amanda A. and {Kim}, Jaeyeon and {Klessen}, Ralf S. and {Kreckel}, Kathryn and {Koch}, Eric W. and {Kruijssen}, J.~M. Diederik and {Larson}, Kirsten L. and {Lee}, Janice C. and {Lopez}, Laura A. and {Machado}, Josh and {Mayker}, Ness and {McElroy}, Rebecca and {Murphy}, Eric J. and {Ostriker}, Eve C. and {Pan}, Hsi-An and {Pessa}, Ismael and {Puschnig}, Johannes and {Razza}, Alessandro and {S{\'a}nchez-Bl{\'a}zquez}, Patricia and {Santoro}, Francesco and {Sardone}, Amy and {Scheuermann}, Fabian and {Sliwa}, Kazimierz and {Sormani}, Mattia C. and {Stuber}, Sophia K. and {Thilker}, David A. and {Turner}, Jordan A. and {Utomo}, Dyas and {Watkins}, Elizabeth J. and {Whitmore}, Bradley},
        title = "{PHANGS-ALMA: Arcsecond CO(2-1) Imaging of Nearby Star-forming Galaxies}",
      journal = {\apjs},
         year = 2021,
        month = dec,
       volume = {257},
       number = {2},
          eid = {43},
        pages = {43},
          doi = {10.3847/1538-4365/ac17f3},
archivePrefix = {arXiv},
       eprint = {2104.07739},
 primaryClass = {astro-ph.GA},
       adsurl = {https://ui.adsabs.harvard.edu/abs/2021ApJS..257...43L}
}

@ARTICLE{QuerejetaETAL2021,
       author = {{Querejeta}, M. and {Schinnerer}, E. and {Meidt}, S. and {Sun}, J. and {Leroy}, A.~K. and {Emsellem}, E. and {Klessen}, R.~S. and {Mu{\~n}oz-Mateos}, J.~C. and {Salo}, H. and {Laurikainen}, E. and {Be{\v{s}}li{\'c}}, I. and {Blanc}, G.~A. and {Chevance}, M. and {Dale}, D.~A. and {Eibensteiner}, C. and {Faesi}, C. and {Garc{\'\i}a-Rodr{\'\i}guez}, A. and {Glover}, S.~C.~O. and {Grasha}, K. and {Henshaw}, J. and {Herrera}, C. and {Hughes}, A. and {Kreckel}, K. and {Kruijssen}, J.~M.~D. and {Liu}, D. and {Murphy}, E.~J. and {Pan}, H. -A. and {Pety}, J. and {Razza}, A. and {Rosolowsky}, E. and {Saito}, T. and {Schruba}, A. and {Usero}, A. and {Watkins}, E.~J. and {Williams}, T.~G.},
        title = "{Stellar structures, molecular gas, and star formation across the PHANGS sample of nearby galaxies}",
      journal = {\aap},
         year = 2021,
        month = dec,
       volume = {656},
          eid = {A133},
        pages = {A133},
          doi = {10.1051/0004-6361/202140695},
archivePrefix = {arXiv},
       eprint = {2109.04491},
 primaryClass = {astro-ph.GA},
       adsurl = {https://ui.adsabs.harvard.edu/abs/2021A&A...656A.133Q}
}

@ARTICLE{PetyETAL2013,
       author = {{Pety}, J{\'e}r{\^o}me and {Schinnerer}, Eva and {Leroy}, Adam K. and {Hughes}, Annie and {Meidt}, Sharon E. and {Colombo}, Dario and {Dumas}, Gaelle and {Garc{\'\i}a-Burillo}, Santiago and {Schuster}, Karl F. and {Kramer}, Carsten and {Dobbs}, Clare L. and {Thompson}, Todd A.},
        title = "{The Plateau de Bure + 30 m Arcsecond Whirlpool Survey Reveals a Thick Disk of Diffuse Molecular Gas in the M51 Galaxy}",
      journal = {\apj},
         year = 2013,
        month = dec,
       volume = {779},
       number = {1},
          eid = {43},
        pages = {43},
          doi = {10.1088/0004-637X/779/1/43},
archivePrefix = {arXiv},
       eprint = {1304.1396},
 primaryClass = {astro-ph.GA},
       adsurl = {https://ui.adsabs.harvard.edu/abs/2013ApJ...779...43P}
}

@ARTICLE{FujimotoETAL2020b,
       author = {{Fujimoto}, Yusuke and {Krumholz}, Mark R. and {Inutsuka}, Shu-ichiro},
        title = "{Distribution and kinematics of $^{26}$Al in the Galactic disc}",
      journal = {\mnras},
         year = 2020,
        month = sep,
       volume = {497},
       number = {2},
        pages = {2442-2454},
          doi = {10.1093/mnras/staa2125},
archivePrefix = {arXiv},
       eprint = {2006.03057},
 primaryClass = {astro-ph.GA},
       adsurl = {https://ui.adsabs.harvard.edu/abs/2020MNRAS.497.2442F}
}

@ARTICLE{FujimotoETAL2020,
       author = {{Fujimoto}, Yusuke and {Maeda}, Fumiya and {Habe}, Asao and {Ohta}, Kouji},
        title = "{Fast cloud-cloud collisions in a strongly barred galaxy: suppression of massive star formation}",
      journal = {\mnras},
         year = 2020,
        month = may,
       volume = {494},
       number = {2},
        pages = {2131-2146},
          doi = {10.1093/mnras/staa840},
archivePrefix = {arXiv},
       eprint = {2003.12074},
 primaryClass = {astro-ph.GA},
       adsurl = {https://ui.adsabs.harvard.edu/abs/2020MNRAS.494.2131F}
}

@ARTICLE{MaedaETAL2020,
       author = {{Maeda}, Fumiya and {Ohta}, Kouji and {Fujimoto}, Yusuke and {Habe}, Asao},
        title = "{Properties of giant molecular clouds in the strongly barred galaxy NGC 1300}",
      journal = {\mnras},
         year = 2020,
        month = apr,
       volume = {493},
       number = {4},
        pages = {5045-5061},
          doi = {10.1093/mnras/staa556},
archivePrefix = {arXiv},
       eprint = {2002.08977},
 primaryClass = {astro-ph.GA},
       adsurl = {https://ui.adsabs.harvard.edu/abs/2020MNRAS.493.5045M}
}

@ARTICLE{SolomonETAL1987,
       author = {{Solomon}, P.~M. and {Rivolo}, A.~R. and {Barrett}, J. and {Yahil}, A.},
        title = "{Mass, Luminosity, and Line Width Relations of Galactic Molecular Clouds}",
      journal = {\apj},
         year = 1987,
        month = aug,
       volume = {319},
        pages = {730},
          doi = {10.1086/165493},
       adsurl = {https://ui.adsabs.harvard.edu/abs/1987ApJ...319..730S}
}

@ARTICLE{Heyer&Dame2015,
       author = {{Heyer}, Mark and {Dame}, T.~M.},
        title = "{Molecular Clouds in the Milky Way}",
      journal = {\araa},
         year = 2015,
        month = aug,
       volume = {53},
        pages = {583-629},
          doi = {10.1146/annurev-astro-082214-122324},
       adsurl = {https://ui.adsabs.harvard.edu/abs/2015ARA&A..53..583H}
}

@ARTICLE{ColomboETAL2022,
       author = {{Colombo}, D. and {Duarte-Cabral}, A. and {Pettitt}, A.~R. and {Urquhart}, J.~S. and {Wyrowski}, F. and {Csengeri}, T. and {Neralwar}, K.~R. and {Schuller}, F. and {Menten}, K.~M. and {Anderson}, L. and {Barnes}, P. and {Beuther}, H. and {Bronfman}, L. and {Eden}, D. and {Ginsburg}, A. and {Henning}, T. and {K{\"o}nig}, C. and {Lee}, M. -Y. and {Mattern}, M. and {Medina}, S. and {Ragan}, S.~E. and {Rigby}, A.~J. and {S{\'a}nchez-Monge}, {\'A}. and {Traficante}, A. and {Yang}, A.~Y. and {Wienen}, M.},
        title = "{The SEDIGISM survey: The influence of spiral arms on the molecular gas distribution of the inner Milky Way}",
      journal = {\aap},
         year = 2022,
        month = feb,
       volume = {658},
          eid = {A54},
        pages = {A54},
          doi = {10.1051/0004-6361/202141287},
archivePrefix = {arXiv},
       eprint = {2110.06071},
 primaryClass = {astro-ph.GA},
       adsurl = {https://ui.adsabs.harvard.edu/abs/2022A&A...658A..54C}
}

@ARTICLE{Roman-DuvalETAL2010,
       author = {{Roman-Duval}, Julia and {Jackson}, James M. and {Heyer}, Mark and {Rathborne}, Jill and {Simon}, Robert},
        title = "{Physical Properties and Galactic Distribution of Molecular Clouds Identified in the Galactic Ring Survey}",
      journal = {\apj},
         year = 2010,
        month = nov,
       volume = {723},
       number = {1},
        pages = {492-507},
          doi = {10.1088/0004-637X/723/1/492},
archivePrefix = {arXiv},
       eprint = {1010.2798},
 primaryClass = {astro-ph.GA},
       adsurl = {https://ui.adsabs.harvard.edu/abs/2010ApJ...723..492R}
}

@InProceedings{Scoville&Sanders1987,
    author="Scoville, N. Z.
    and Sanders, D. B.",
    editor="Hollenbach, David J.
    and Thronson, Harley A.",
    title="H2 in the Galaxy",
    booktitle="Interstellar Processes",
    year="1987",
    publisher="Springer Netherlands",
    address="Dordrecht",
    pages="21--50",
    isbn="978-94-009-3861-8"
}

@ARTICLE{ChurchwellETAL2009,
       author = {{Churchwell}, Ed and {Babler}, Brian L. and {Meade}, Marilyn R. and {Whitney}, Barbara A. and {Benjamin}, Robert and {Indebetouw}, Remy and {Cyganowski}, Claudia and {Robitaille}, Thomas P. and {Povich}, Matthew and {Watson}, Christer and {Bracker}, Steve},
        title = "{The Spitzer/GLIMPSE Surveys: A New View of the Milky Way}",
      journal = {\pasp},
         year = 2009,
        month = mar,
       volume = {121},
       number = {877},
        pages = {213},
          doi = {10.1086/597811},
       adsurl = {https://ui.adsabs.harvard.edu/abs/2009PASP..121..213C}
}

@ARTICLE{SawadaETAL2018,
       author = {{Sawada}, Tsuyoshi and {Koda}, Jin and {Hasegawa}, Tetsuo},
        title = "{Internal Structures of Molecular Clouds in the LMC Revealed by ALMA}",
      journal = {\apj},
         year = 2018,
        month = nov,
       volume = {867},
       number = {2},
          eid = {166},
        pages = {166},
          doi = {10.3847/1538-4357/aae395},
archivePrefix = {arXiv},
       eprint = {1812.06001},
 primaryClass = {astro-ph.GA},
       adsurl = {https://ui.adsabs.harvard.edu/abs/2018ApJ...867..166S}
}

@ARTICLE{ReidETAL2016,
       author = {{Reid}, M.~J. and {Dame}, T.~M. and {Menten}, K.~M. and {Brunthaler}, A.},
        title = "{A Parallax-based Distance Estimator for Spiral Arm Sources}",
      journal = {\apj},
         year = 2016,
        month = jun,
       volume = {823},
       number = {2},
          eid = {77},
        pages = {77},
          doi = {10.3847/0004-637X/823/2/77},
archivePrefix = {arXiv},
       eprint = {1604.02433},
 primaryClass = {astro-ph.GA},
       adsurl = {https://ui.adsabs.harvard.edu/abs/2016ApJ...823...77R}
}

@ARTICLE{EgusaETAL2018,
       author = {{Egusa}, Fumi and {Hirota}, Akihiko and {Baba}, Junichi and {Muraoka}, Kazuyuki},
        title = "{Molecular Gas Properties in M83 from CO PDFs}",
      journal = {\apj},
         year = 2018,
        month = feb,
       volume = {854},
       number = {2},
          eid = {90},
        pages = {90},
          doi = {10.3847/1538-4357/aaa76d},
archivePrefix = {arXiv},
       eprint = {1801.04025},
 primaryClass = {astro-ph.GA},
       adsurl = {https://ui.adsabs.harvard.edu/abs/2018ApJ...854...90E}
}

@ARTICLE{ColomboETAL2014,
       author = {{Colombo}, Dario and {Hughes}, Annie and {Schinnerer}, Eva and
         {Meidt}, Sharon E. and {Leroy}, Adam K. and {Pety}, J{\'e}r{\^o}me and
         {Dobbs}, Clare L. and {Garc{\'\i}a-Burillo}, Santiago and
         {Dumas}, Ga{\"e}lle and {Thompson}, Todd A. and {Schuster}, Karl F. and
         {Kramer}, Carsten},
        title = "{The PdBI Arcsecond Whirlpool Survey (PAWS): Environmental Dependence of Giant Molecular Cloud Properties in M51}",
      journal = {\apj},
         year = 2014,
        month = mar,
       volume = {784},
       number = {1},
          eid = {3},
        pages = {3},
          doi = {10.1088/0004-637X/784/1/3},
archivePrefix = {arXiv},
       eprint = {1401.1505},
 primaryClass = {astro-ph.GA},
       adsurl = {https://ui.adsabs.harvard.edu/abs/2014ApJ...784....3C}
}

@ARTICLE{BigielETAL2008,
       author = {{Bigiel}, F. and {Leroy}, A. and {Walter}, F. and {Brinks}, E. and {de Blok}, W.~J.~G. and {Madore}, B. and {Thornley}, M.~D.},
        title = "{The Star Formation Law in Nearby Galaxies on Sub-Kpc Scales}",
      journal = {\aj},
         year = 2008,
        month = dec,
       volume = {136},
       number = {6},
        pages = {2846-2871},
          doi = {10.1088/0004-6256/136/6/2846},
archivePrefix = {arXiv},
       eprint = {0810.2541},
 primaryClass = {astro-ph},
       adsurl = {https://ui.adsabs.harvard.edu/abs/2008AJ....136.2846B}
}

@ARTICLE{CalzettiETAL2007,
       author = {{Calzetti}, D. and {Kennicutt}, R.~C. and {Engelbracht}, C.~W. and {Leitherer}, C. and {Draine}, B.~T. and {Kewley}, L. and {Moustakas}, J. and {Sosey}, M. and {Dale}, D.~A. and {Gordon}, K.~D. and {Helou}, G.~X. and {Hollenbach}, D.~J. and {Armus}, L. and {Bendo}, G. and {Bot}, C. and {Buckalew}, B. and {Jarrett}, T. and {Li}, A. and {Meyer}, M. and {Murphy}, E.~J. and {Prescott}, M. and {Regan}, M.~W. and {Rieke}, G.~H. and {Roussel}, H. and {Sheth}, K. and {Smith}, J.~D.~T. and {Thornley}, M.~D. and {Walter}, F.},
        title = "{The Calibration of Mid-Infrared Star Formation Rate Indicators}",
      journal = {\apj},
         year = 2007,
        month = sep,
       volume = {666},
       number = {2},
        pages = {870-895},
          doi = {10.1086/520082},
archivePrefix = {arXiv},
       eprint = {0705.3377},
 primaryClass = {astro-ph},
       adsurl = {https://ui.adsabs.harvard.edu/abs/2007ApJ...666..870C}
}

@ARTICLE{ThimETAL2003,
       author = {{Thim}, Frank and {Tammann}, G.~A. and {Saha}, A. and {Dolphin}, A. and {Sandage}, Allan and {Tolstoy}, E. and {Labhardt}, Lukas},
        title = "{The Cepheid Distance to NGC 5236 (M83) with the ESO Very Large Telescope}",
      journal = {\apj},
         year = 2003,
        month = jun,
       volume = {590},
       number = {1},
        pages = {256-270},
          doi = {10.1086/374888},
archivePrefix = {arXiv},
       eprint = {astro-ph/0303101},
 primaryClass = {astro-ph},
       adsurl = {https://ui.adsabs.harvard.edu/abs/2003ApJ...590..256T}
}

@ARTICLE{ThatteETAL2000,
       author = {{Thatte}, N. and {Tecza}, M. and {Genzel}, R.},
        title = "{Stellar dynamics observations of a double nucleus in M 83}",
      journal = {\aap},
         year = 2000,
        month = dec,
       volume = {364},
        pages = {L47-L53},
          doi = {10.48550/arXiv.astro-ph/0009392},
archivePrefix = {arXiv},
       eprint = {astro-ph/0009392},
 primaryClass = {astro-ph},
       adsurl = {https://ui.adsabs.harvard.edu/abs/2000A&A...364L..47T}
}

@BOOK{deVaucouleursETAL1991book,
       author = {{de Vaucouleurs}, Gerard and {de Vaucouleurs}, Antoinette and {Corwin}, Herold G., Jr. and {Buta}, Ronald J. and {Paturel}, Georges and {Fouque}, Pascal},
        title = "{Third Reference Catalogue of Bright Galaxies}",
         year = 1991,
       adsurl = {https://ui.adsabs.harvard.edu/abs/1991rc3..book.....D}
}

@ARTICLE{HandaETAL1990,
       author = {{Handa}, Toshihiro and {Nakai}, Naomasa and {Sofue}, Yoshiaki and {Hayashi}, Masahiko and {Fujimoto}, Mitsuaki},
        title = "{CO Line Observations of the Bar and Nucleus of the Barred Spiral Galaxy M83}",
      journal = {\pasj},
         year = 1990,
        month = feb,
       volume = {42},
        pages = {1-17},
       adsurl = {https://ui.adsabs.harvard.edu/abs/1990PASJ...42....1H}
}

@ARTICLE{SawadaETAL2012a,
       author = {{Sawada}, Tsuyoshi and {Hasegawa}, Tetsuo and {Sugimoto}, Masahiro and {Koda}, Jin and {Handa}, Toshihiro},
        title = "{Structural Variation of Molecular Gas in the Sagittarius Arm and Interarm Regions}",
      journal = {\apj},
         year = 2012,
        month = jun,
       volume = {752},
       number = {2},
          eid = {118},
        pages = {118},
          doi = {10.1088/0004-637X/752/2/118},
archivePrefix = {arXiv},
       eprint = {1204.5687},
 primaryClass = {astro-ph.GA},
       adsurl = {https://ui.adsabs.harvard.edu/abs/2012ApJ...752..118S}
}

@ARTICLE{SawadaETAL2012b,
       author = {{Sawada}, Tsuyoshi and {Hasegawa}, Tetsuo and {Koda}, Jin},
        title = "{Structured Molecular Gas Reveals Galactic Spiral Arms}",
      journal = {\apjl},
         year = 2012,
        month = nov,
       volume = {759},
       number = {1},
          eid = {L26},
        pages = {L26},
          doi = {10.1088/2041-8205/759/1/L26},
archivePrefix = {arXiv},
       eprint = {1210.1520},
 primaryClass = {astro-ph.GA},
       adsurl = {https://ui.adsabs.harvard.edu/abs/2012ApJ...759L..26S}
}

@ARTICLE{SchneiderETAL2022,
       author = {{Schneider}, N. and {Ossenkopf-Okada}, V. and {Clarke}, S. and {Klessen}, R.~S. and {Kabanovic}, S. and {Veltchev}, T. and {Bontemps}, S. and {Dib}, S. and {Csengeri}, T. and {Federrath}, C. and {Di Francesco}, J. and {Motte}, F. and {Andr{\'e}}, Ph. and {Arzoumanian}, D. and {Beattie}, J.~R. and {Bonne}, L. and {Didelon}, P. and {Elia}, D. and {K{\"o}nyves}, V. and {Kritsuk}, A. and {Ladjelate}, B. and {Myers}, Ph. and {Pezzuto}, S. and {Robitaille}, J.~F. and {Roy}, A. and {Seifried}, D. and {Simon}, R. and {Soler}, J. and {Ward-Thompson}, D.},
        title = "{Understanding star formation in molecular clouds. IV. Column density PDFs from quiescent to massive molecular clouds}",
      journal = {\aap},
         year = 2022,
        month = oct,
       volume = {666},
          eid = {A165},
        pages = {A165},
          doi = {10.1051/0004-6361/202039610},
archivePrefix = {arXiv},
       eprint = {2207.14604},
 primaryClass = {astro-ph.GA},
       adsurl = {https://ui.adsabs.harvard.edu/abs/2022A&A...666A.165S}
}

@ARTICLE{BurkhartETALETAL2015,
       author = {{Burkhart}, Blakesley and {Collins}, David C. and {Lazarian}, Alex},
        title = "{Observational Diagnostics of Self-gravitating MHD Turbulence in Giant Molecular Clouds}",
      journal = {\apj},
         year = 2015,
        month = jul,
       volume = {808},
       number = {1},
          eid = {48},
        pages = {48},
          doi = {10.1088/0004-637X/808/1/48},
archivePrefix = {arXiv},
       eprint = {1505.03855},
 primaryClass = {astro-ph.SR},
       adsurl = {https://ui.adsabs.harvard.edu/abs/2015ApJ...808...48B}
}

@ARTICLE{Federrath&Klessen2013,
       author = {{Federrath}, Christoph and {Klessen}, Ralf S.},
        title = "{On the Star Formation Efficiency of Turbulent Magnetized Clouds}",
      journal = {\apj},
         year = 2013,
        month = jan,
       volume = {763},
       number = {1},
          eid = {51},
        pages = {51},
          doi = {10.1088/0004-637X/763/1/51},
archivePrefix = {arXiv},
       eprint = {1211.6433},
 primaryClass = {astro-ph.SR},
       adsurl = {https://ui.adsabs.harvard.edu/abs/2013ApJ...763...51F}
}

@ARTICLE{MatsusakaETAL2024,
       author = {{Matsusaka}, Ren and {Handa}, Toshihiro and {Fujimoto}, Yusuke and {Murase}, Takeru and {Hirata}, Yushi and {Nishi}, Junya and {Ito}, Takumi and {Sasaki}, Megumi and {Mizoguchi}, Tomoki},
        title = "{Sub-kpc scale gas density histogram of the galactic molecular gas: a new statistical method to characterize galactic-scale gas structures}",
      journal = {\mnras},
         year = 2024,
        month = feb,
       volume = {528},
       number = {2},
        pages = {3473-3485},
          doi = {10.1093/mnras/stae227},
archivePrefix = {arXiv},
       eprint = {2402.02821},
 primaryClass = {astro-ph.GA},
       adsurl = {https://ui.adsabs.harvard.edu/abs/2024MNRAS.528.3473M}
}

@ARTICLE{RosolowskyETAL2021,
       author = {{Rosolowsky}, Erik and {Hughes}, Annie and {Leroy}, Adam K. and {Sun}, Jiayi and {Querejeta}, Miguel and {Schruba}, Andreas and {Usero}, Antonio and {Herrera}, Cinthya N. and {Liu}, Daizhong and {Pety}, J{\'e}r{\^o}me and {Saito}, Toshiki and {Be{\v{s}}li{\'c}}, Ivana and {Bigiel}, Frank and {Blanc}, Guillermo and {Chevance}, M{\'e}lanie and {Dale}, Daniel A. and {Deger}, Sinan and {Faesi}, Christopher M. and {Glover}, Simon C.~O. and {Henshaw}, Jonathan D. and {Klessen}, Ralf S. and {Kruijssen}, J.~M. Diederik and {Larson}, Kirsten and {Lee}, Janice and {Meidt}, Sharon and {Mok}, Angus and {Schinnerer}, Eva and {Thilker}, David A. and {Williams}, Thomas G.},
        title = "{Giant molecular cloud catalogues for PHANGS-ALMA: methods and initial results}",
      journal = {\mnras},
         year = 2021,
        month = mar,
       volume = {502},
       number = {1},
        pages = {1218-1245},
          doi = {10.1093/mnras/stab085},
archivePrefix = {arXiv},
       eprint = {2101.04697},
 primaryClass = {astro-ph.GA},
       adsurl = {https://ui.adsabs.harvard.edu/abs/2021MNRAS.502.1218R}
}

@ARTICLE{SunETAL2018,
       author = {{Sun}, Jiayi and {Leroy}, Adam K. and {Schruba}, Andreas and {Rosolowsky}, Erik and {Hughes}, Annie and {Kruijssen}, J.~M. Diederik and {Meidt}, Sharon and {Schinnerer}, Eva and {Blanc}, Guillermo A. and {Bigiel}, Frank and {Bolatto}, Alberto D. and {Chevance}, M{\'e}lanie and {Groves}, Brent and {Herrera}, Cinthya N. and {Hygate}, Alexander P.~S. and {Pety}, J{\'e}r{\^o}me and {Querejeta}, Miguel and {Usero}, Antonio and {Utomo}, Dyas},
        title = "{Cloud-scale Molecular Gas Properties in 15 Nearby Galaxies}",
      journal = {\apj},
         year = 2018,
        month = jun,
       volume = {860},
       number = {2},
          eid = {172},
        pages = {172},
          doi = {10.3847/1538-4357/aac326},
archivePrefix = {arXiv},
       eprint = {1805.00937},
 primaryClass = {astro-ph.GA},
       adsurl = {https://ui.adsabs.harvard.edu/abs/2018ApJ...860..172S}
}

@ARTICLE{KodaETAL2019,
       author = {{Koda}, Jin and {Teuben}, Peter and {Sawada}, Tsuyoshi and {Plunkett}, Adele and {Fomalont}, Ed},
        title = "{Total Power Map to Visibilities (TP2VIS): Joint Deconvolution of ALMA 12m, 7m, and Total Power Array Data}",
      journal = {\pasp},
         year = 2019,
        month = may,
       volume = {131},
       number = {999},
        pages = {054505},
          doi = {10.1088/1538-3873/ab047e},
archivePrefix = {arXiv},
       eprint = {1903.07611},
 primaryClass = {astro-ph.IM},
       adsurl = {https://ui.adsabs.harvard.edu/abs/2019PASP..131e4505K}
}

@INPROCEEDINGS{SaultETAL1995,
       author = {{Sault}, R.~J. and {Teuben}, P.~J. and {Wright}, M.~C.~H.},
        title = "{A Retrospective View of MIRIAD}",
    booktitle = {Astronomical Data Analysis Software and Systems IV},
         year = 1995,
       editor = {{Shaw}, R.~A. and {Payne}, H.~E. and {Hayes}, J.~J.~E.},
       series = {Astronomical Society of the Pacific Conference Series},
       volume = {77},
        month = jan,
        pages = {433},
          doi = {10.48550/arXiv.astro-ph/0612759},
archivePrefix = {arXiv},
       eprint = {astro-ph/0612759},
 primaryClass = {astro-ph},
       adsurl = {https://ui.adsabs.harvard.edu/abs/1995ASPC...77..433S}
}

@ARTICLE{SaultETAL1996,
       author = {{Sault}, R.~J. and {Staveley-Smith}, L. and {Brouw}, W.~N.},
        title = "{An approach to interferometric mosaicing.}",
      journal = {\aaps},
         year = 1996,
        month = dec,
       volume = {120},
        pages = {375-384},
       adsurl = {https://ui.adsabs.harvard.edu/abs/1996A&AS..120..375S}
}

@ARTICLE{KodaETAL2023,
       author = {{Koda}, Jin and {Hirota}, Akihiko and {Egusa}, Fumi and {Sakamoto}, Kazushi and {Sawada}, Tsuyoshi and {Heyer}, Mark and {Baba}, Junichi and {Boissier}, Samuel and {Calzetti}, Daniela and {Meyer}, Jennifer Donovan and {Elmegreen}, Bruce G. and {de Paz}, Armando Gil and {Harada}, Nanase and {Ho}, Luis C. and {Kobayashi}, Masato I.~N. and {Kuno}, Nario and {Lee}, Amanda M. and {Madore}, Barry F. and {Maeda}, Fumiya and {Mart{\'\i}n}, Sergio and {Muraoka}, Kazuyuki and {Nakanishi}, Kouichiro and {Onodera}, Sachiko and {Pineda}, Jorge L. and {Scoville}, Nick and {Watanabe}, Yoshimasa},
        title = "{Diverse Molecular Structures across the Whole Star-forming Disk of M83: High-fidelity Imaging at 40 pc Resolution}",
      journal = {\apj},
         year = 2023,
        month = jun,
       volume = {949},
       number = {2},
          eid = {108},
        pages = {108},
          doi = {10.3847/1538-4357/acc65e},
archivePrefix = {arXiv},
       eprint = {2303.12108},
 primaryClass = {astro-ph.GA},
       adsurl = {https://ui.adsabs.harvard.edu/abs/2023ApJ...949..108K}
}

@ARTICLE{Rosolowsky2007,
       author = {{Rosolowsky}, Erik and {Keto}, Eric and {Matsushita}, Satoki and {Willner}, S.~P.},
        title = "{High-Resolution Molecular Gas Maps of M33}",
      journal = {\apj},
         year = 2007,
        month = jun,
       volume = {661},
       number = {2},
        pages = {830-844},
          doi = {10.1086/516621},
archivePrefix = {arXiv},
       eprint = {astro-ph/0703006},
 primaryClass = {astro-ph},
       adsurl = {https://ui.adsabs.harvard.edu/abs/2007ApJ...661..830R}
}

@ARTICLE{DemachiETAL2024,
       author = {{Demachi}, Fumika and {Fukui}, Yasuo and {Yamada}, Rin I. and {Tachihara}, Kengo and {Hayakawa}, Takahiro and {Tokuda}, Kazuki and {Fujita}, Shinji and {Kobayashi}, Masato I.~N. and {Muraoka}, Kazuyuki and {Konishi}, Ayu and {Tsuge}, Kisetsu and {Onishi}, Toshikazu and {Kawamura}, Akiko},
        title = "{Giant molecular clouds and their type classification in M 74: Toward understanding star formation and cloud evolution}",
      journal = {\pasj},
         year = 2024,
        month = oct,
       volume = {76},
       number = {5},
        pages = {1059-1083},
          doi = {10.1093/pasj/psae071},
archivePrefix = {arXiv},
       eprint = {2305.19192},
 primaryClass = {astro-ph.GA},
       adsurl = {https://ui.adsabs.harvard.edu/abs/2024PASJ...76.1059D}
}

@ARTICLE{MuraokaETAL2023,
       author = {{Muraoka}, Kazuyuki and {Konishi}, Ayu and {Tokuda}, Kazuki and {Kondo}, Hiroshi and {Miura}, Rie E. and {Tosaki}, Tomoka and {Onodera}, Sachiko and {Kuno}, Nario and {Kobayashi}, Masato I.~N. and {Tsuge}, Kisetsu and {Sano}, Hidetoshi and {Kitano}, Naoya and {Fujita}, Shinji and {Nishimura}, Atsushi and {Onishi}, Toshikazu and {Saigo}, Kazuya and {Yamada}, Rin I. and {Demachi}, Fumika and {Tachihara}, Kengo and {Fukui}, Yasuo and {Kawamura}, Akiko and {AAS Journals Data Editors}},
        title = "{ACA CO(J = 2-1) Mapping of the Nearest Spiral Galaxy M33. I. Initial Results and Identification of Molecular Clouds}",
      journal = {\apj},
         year = 2023,
        month = aug,
       volume = {953},
       number = {2},
          eid = {164},
        pages = {164},
          doi = {10.3847/1538-4357/ace4bd},
archivePrefix = {arXiv},
       eprint = {2307.02039},
 primaryClass = {astro-ph.GA},
       adsurl = {https://ui.adsabs.harvard.edu/abs/2023ApJ...953..164M}
}

@ARTICLE{MuraseETAL2023,
       author = {{Murase}, Takeru and {Handa}, Toshihiro and {Matsusaka}, Ren and {Shimajiri}, Yoshito and {Kobayashi}, Masato I.~N. and {Kohno}, Mikito and {Nishi}, Junya and {Takeba}, Norimi and {Shibata}, Yosuke},
        title = "{Multilognormal density structure in Cygnus-X molecular clouds: a fitting for N-PDF without power law}",
      journal = {\mnras},
         year = 2023,
        month = jul,
       volume = {523},
       number = {1},
        pages = {1373-1387},
          doi = {10.1093/mnras/stad1451},
archivePrefix = {arXiv},
       eprint = {2305.07094},
 primaryClass = {astro-ph.GA},
       adsurl = {https://ui.adsabs.harvard.edu/abs/2023MNRAS.523.1373M}
}

@ARTICLE{astropy_2013,
       author = {{Astropy Collaboration} and {Robitaille}, Thomas P. and {Tollerud}, Erik J. and {Greenfield}, Perry and {Droettboom}, Michael and {Bray}, Erik and {Aldcroft}, Tom and {Davis}, Matt and {Ginsburg}, Adam and {Price-Whelan}, Adrian M. and {Kerzendorf}, Wolfgang E. and {Conley}, Alexander and {Crighton}, Neil and {Barbary}, Kyle and {Muna}, Demitri and {Ferguson}, Henry and {Grollier}, Fr{\'e}d{\'e}ric and {Parikh}, Madhura M. and {Nair}, Prasanth H. and {Unther}, Hans M. and {Deil}, Christoph and {Woillez}, Julien and {Conseil}, Simon and {Kramer}, Roban and {Turner}, James E.~H. and {Singer}, Leo and {Fox}, Ryan and {Weaver}, Benjamin A. and {Zabalza}, Victor and {Edwards}, Zachary I. and {Azalee Bostroem}, K. and {Burke}, D.~J. and {Casey}, Andrew R. and {Crawford}, Steven M. and {Dencheva}, Nadia and {Ely}, Justin and {Jenness}, Tim and {Labrie}, Kathleen and {Lim}, Pey Lian and {Pierfederici}, Francesco and {Pontzen}, Andrew and {Ptak}, Andy and {Refsdal}, Brian and {Servillat}, Mathieu and {Streicher}, Ole},
        title = "{Astropy: A community Python package for astronomy}",
      journal = {\aap},
         year = 2013,
        month = oct,
       volume = {558},
          eid = {A33},
        pages = {A33},
          doi = {10.1051/0004-6361/201322068},
archivePrefix = {arXiv},
       eprint = {1307.6212},
 primaryClass = {astro-ph.IM},
       adsurl = {https://ui.adsabs.harvard.edu/abs/2013A&A...558A..33A}
}

@ARTICLE{astropy_2018,
       author = {{Astropy Collaboration} and {Price-Whelan}, A.~M. and {Sip{\H{o}}cz}, B.~M. and {G{\"u}nther}, H.~M. and {Lim}, P.~L. and {Crawford}, S.~M. and {Conseil}, S. and {Shupe}, D.~L. and {Craig}, M.~W. and {Dencheva}, N. and {Ginsburg}, A. and {VanderPlas}, J.~T. and {Bradley}, L.~D. and {P{\'e}rez-Su{\'a}rez}, D. and {de Val-Borro}, M. and {Aldcroft}, T.~L. and {Cruz}, K.~L. and {Robitaille}, T.~P. and {Tollerud}, E.~J. and {Ardelean}, C. and {Babej}, T. and {Bach}, Y.~P. and {Bachetti}, M. and {Bakanov}, A.~V. and {Bamford}, S.~P. and {Barentsen}, G. and {Barmby}, P. and {Baumbach}, A. and {Berry}, K.~L. and {Biscani}, F. and {Boquien}, M. and {Bostroem}, K.~A. and {Bouma}, L.~G. and {Brammer}, G.~B. and {Bray}, E.~M. and {Breytenbach}, H. and {Buddelmeijer}, H. and {Burke}, D.~J. and {Calderone}, G. and {Cano Rodr{\'\i}guez}, J.~L. and {Cara}, M. and {Cardoso}, J.~V.~M. and {Cheedella}, S. and {Copin}, Y. and {Corrales}, L. and {Crichton}, D. and {D'Avella}, D. and {Deil}, C. and {Depagne}, {\'E}. and {Dietrich}, J.~P. and {Donath}, A. and {Droettboom}, M. and {Earl}, N. and {Erben}, T. and {Fabbro}, S. and {Ferreira}, L.~A. and {Finethy}, T. and {Fox}, R.~T. and {Garrison}, L.~H. and {Gibbons}, S.~L.~J. and {Goldstein}, D.~A. and {Gommers}, R. and {Greco}, J.~P. and {Greenfield}, P. and {Groener}, A.~M. and {Grollier}, F. and {Hagen}, A. and {Hirst}, P. and {Homeier}, D. and {Horton}, A.~J. and {Hosseinzadeh}, G. and {Hu}, L. and {Hunkeler}, J.~S. and {Ivezi{\'c}}, {\v{Z}}. and {Jain}, A. and {Jenness}, T. and {Kanarek}, G. and {Kendrew}, S. and {Kern}, N.~S. and {Kerzendorf}, W.~E. and {Khvalko}, A. and {King}, J. and {Kirkby}, D. and {Kulkarni}, A.~M. and {Kumar}, A. and {Lee}, A. and {Lenz}, D. and {Littlefair}, S.~P. and {Ma}, Z. and {Macleod}, D.~M. and {Mastropietro}, M. and {McCully}, C. and {Montagnac}, S. and {Morris}, B.~M. and {Mueller}, M. and {Mumford}, S.~J. and {Muna}, D. and {Murphy}, N.~A. and {Nelson}, S. and {Nguyen}, G.~H. and {Ninan}, J.~P. and {N{\"o}the}, M. and {Ogaz}, S. and {Oh}, S. and {Parejko}, J.~K. and {Parley}, N. and {Pascual}, S. and {Patil}, R. and {Patil}, A.~A. and {Plunkett}, A.~L. and {Prochaska}, J.~X. and {Rastogi}, T. and {Reddy Janga}, V. and {Sabater}, J. and {Sakurikar}, P. and {Seifert}, M. and {Sherbert}, L.~E. and {Sherwood-Taylor}, H. and {Shih}, A.~Y. and {Sick}, J. and {Silbiger}, M.~T. and {Singanamalla}, S. and {Singer}, L.~P. and {Sladen}, P.~H. and {Sooley}, K.~A. and {Sornarajah}, S. and {Streicher}, O. and {Teuben}, P. and {Thomas}, S.~W. and {Tremblay}, G.~R. and {Turner}, J.~E.~H. and {Terr{\'o}n}, V. and {van Kerkwijk}, M.~H. and {de la Vega}, A. and {Watkins}, L.~L. and {Weaver}, B.~A. and {Whitmore}, J.~B. and {Woillez}, J. and {Zabalza}, V. and {Astropy Contributors}},
        title = "{The Astropy Project: Building an Open-science Project and Status of the v2.0 Core Package}",
      journal = {\aj},
         year = 2018,
        month = sep,
       volume = {156},
       number = {3},
          eid = {123},
        pages = {123},
          doi = {10.3847/1538-3881/aabc4f},
archivePrefix = {arXiv},
       eprint = {1801.02634},
 primaryClass = {astro-ph.IM},
       adsurl = {https://ui.adsabs.harvard.edu/abs/2018AJ....156..123A}
}

@Article{Hunter2007,
  Author    = {Hunter, J. D.},
  Title     = {Matplotlib: A 2D graphics environment},
  Journal   = {Computing in Science \& Engineering},
  Volume    = {9},
  Number    = {3},
  Pages     = {90--95},
  publisher = {IEEE COMPUTER SOC},
  doi       = {10.1109/MCSE.2007.55},
  year      = 2007
}

@Article{HarrisETAL2020,
 title         = {Array programming with {NumPy}},
 author        = {Charles R. Harris and K. Jarrod Millman and St{\'{e}}fan J.
                 van der Walt and Ralf Gommers and Pauli Virtanen and David
                 Cournapeau and Eric Wieser and Julian Taylor and Sebastian
                 Berg and Nathaniel J. Smith and Robert Kern and Matti Picus
                 and Stephan Hoyer and Marten H. van Kerkwijk and Matthew
                 Brett and Allan Haldane and Jaime Fern{\'{a}}ndez del
                 R{\'{i}}o and Mark Wiebe and Pearu Peterson and Pierre
                 G{\'{e}}rard-Marchant and Kevin Sheppard and Tyler Reddy and
                 Warren Weckesser and Hameer Abbasi and Christoph Gohlke and
                 Travis E. Oliphant},
 year          = {2020},
 month         = sep,
 journal       = {Nature},
 volume        = {585},
 number        = {7825},
 pages         = {357--362},
 doi           = {10.1038/s41586-020-2649-2},
 publisher     = {Springer Science and Business Media {LLC}},
 url           = {https://doi.org/10.1038/s41586-020-2649-2}
}




\appendix
\section{Comparison with the Milky Way definition of the H-LN fraction}
\label{app:HLN-fraction}

\begin{figure}
    \includegraphics[width=\linewidth,trim={-15 25 10 25}, clip]{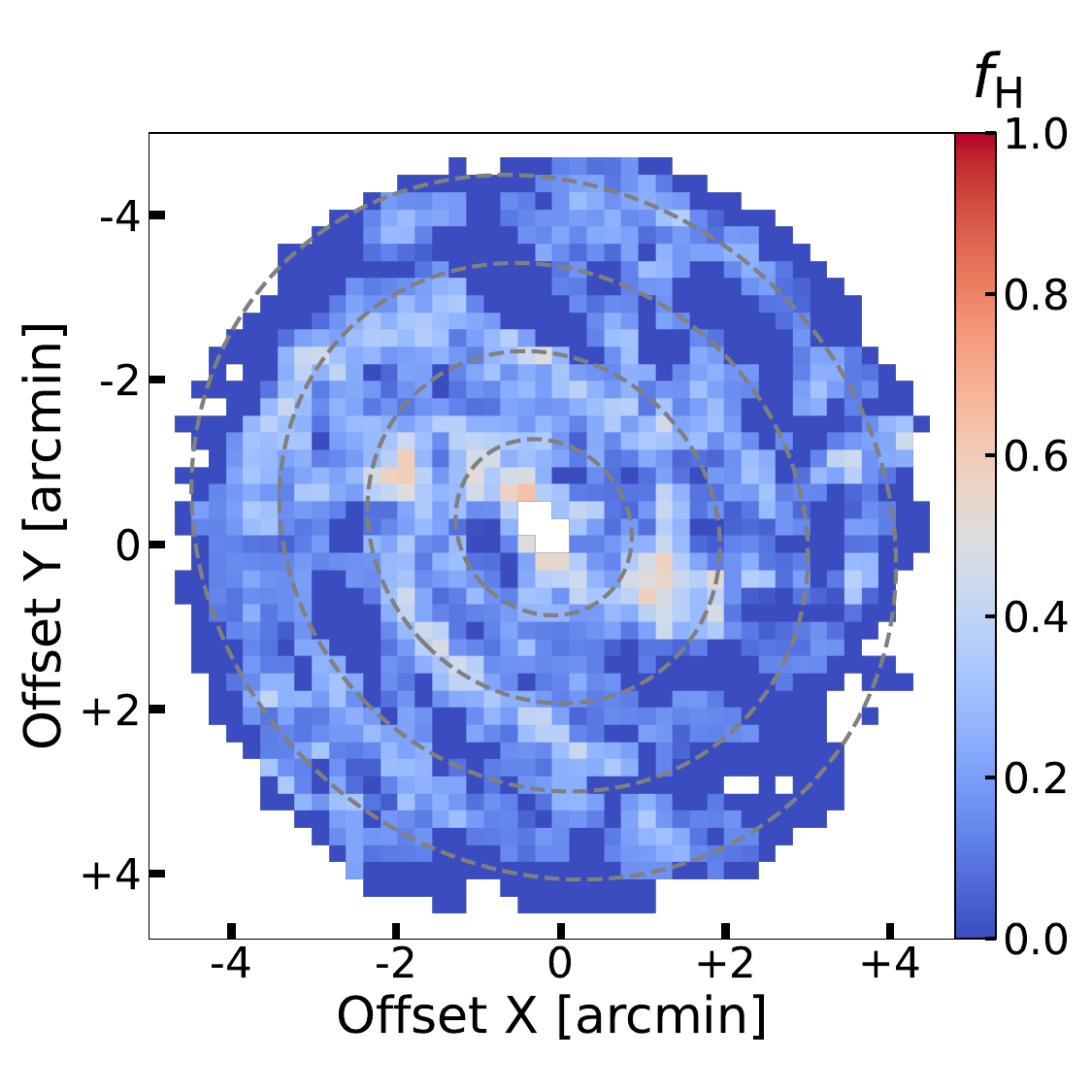}
    \caption{Spatial distribution of the area-based H-LN fraction, $f_{\rm H}$, in M83, calculated using the same definition as in \citet{MatsusakaETAL2024}. The colour scale and range are the same as Fig.~\ref{fig:spatial_dist}c and d.}
    \label{fig:fh_Matsusaka+24}
\end{figure}

For direct comparison with \citet{MatsusakaETAL2024}, we also calculated the H-LN fraction for M83 using the same zeroth-moment-based definition:
\begin{equation}
f_{\rm H}=
\frac{\int N_{\rm H}(\Sigma_{\rm mol})\, d\Sigma_{\rm mol}}
{\int N_{\rm L}(\Sigma_{\rm mol})\, d\Sigma_{\rm mol}
+\int N_{\rm H}(\Sigma_{\rm mol})\, d\Sigma_{\rm mol}}.
\end{equation}
As in the main analysis, we set $f_{\rm H}=0$ for GDH cells in which the H-LN component is not identified. Fig.~\ref{fig:fh_Matsusaka+24} shows the spatial distribution of $f_{\rm H}$ in M83. This trend is similar to that found in the Milky Way, spanning a comparable range from $\sim 0.2$ to $\sim 0.6$ in both galaxies. This similarity supports the applicability of the GDH methodology across different galactic systems. Regions with high $f_{\rm H}$ form coherent structures \citep{MatsusakaETAL2024}, and in M83 these structures can be directly associated with the spiral arms and the bar end thanks to its nearly face-on view.



\bsp	
\label{lastpage}
\end{document}